\documentclass{aastex631}

\graphicspath{{./}{images/}}

\begin{document}

\title{Test particle energization of heavy ions in magnetohydrodynamic turbulence}

\author[0000-0002-5276-9351]{F. Pugliese}
\affiliation{Departamento de Fisica, Facultad de Ciencias Exactas y Naturales,\\
Universidad de Buenos Aires e Instituto de Fisica de Buenos Aires (IFIBA, CONICET), Buenos Aires, Argentina}

\author{P. Dmitruk}
\affiliation{Departamento de Fisica, Facultad de Ciencias Exactas y Naturales,\\
Universidad de Buenos Aires e Instituto de Fisica de Buenos Aires (IFIBA, CONICET), Buenos Aires, Argentina}




\begin{abstract}

In the present work, we study the energization and displacement of heavy ions through the use of test particles interacting with the electromagnetic fields of magnetohydrodynamic (MHD) turbulence.
These fields are obtained from pseudospectral direct numerical solutions (DNSs) of the compressible three-dimensional MHD equations with a strong background magnetic field.
We find particle energization to be predominantly perpendicular as the ions become heavier (lower charge-to-mass ratio) and that high displacement is detrimental for perpendicular energization.
On the other hand, perpendicular displacement is unaffected by the charge-to-mass ratio, which we explain with a simple guide center model.
Using Voronoi tessellation along with this model, we analyze preferential concentration and find that particles behave as tracers in the perpendicular plane, clustering in regions with $\nabla_\perp\cdot\mathbf{u}_\perp < 0$.
These regions also have $(\nabla\times\mathbf{E})_z < 0$, which is optimal for perpendicular energization, thus providing a mechanism to understand precedent results.

\end{abstract}



\section{Introduction}{\label{sec:intro}}

In the Universe, baryonic matter is found almost exclusively in the form of plasma.
Particle energization arises from energy conversion between electromagnetic, kinetic, thermal, and non-thermal energies in these plasmas (\cite{Retino2021}).
Pickup ions (PUIs) are singly-charged particles present in the heliosphere originated from the ionization of neutral material (mainly atoms but also molecules) (\cite{Gloeckler2001}).
This ionization may be due to charge exchange, electron impact or photoionization and more mechanisms are still being proposed (\cite{Quinn2018}).
The first detection of He$^+$ at 1AU with AMPTE (\cite{Mobius1985}) was followed by multiple more (\cite{Burlaga1996, McComas2017}).
Observations by the SWICS (\cite{Gloeckler1992}) in \textit{Ulysses} also found the presence of H$^+$, He$^+$, C$^+$, N$^+$, O$^+$ and Ne$^+$ (\cite{Gloeckler1998}), with different temperatures for each species (\cite{Hefti1998, Gershman2012}).
Ion thermalization is achieved through collisions with protons, but this equilibration time scales as $m/q^2$(\cite{Spitzer1963}) and tends to be higher than the expansion time of the solar wind.
Thus, thermalization is slow and heavy ions have higher thermal velocities than protons, depending mostly on their charge-to-mass ratio; in \cite{Tracy2015} He$^{2+}$ and C$^{6+}$ where found to have similar thermal velocities and up to 50\% higher than protons.

PUIs are found to reach up to twice the solar wind speed, but even higher energies may be achieved by those produced by solar particle events (\cite{Vlahos2019}).
These highly energetic ions can cause extreme biological damage to cells and tissue, resulting in health problems ranging from radiation sickness (in the short-term) to cancer (in the long-term).
Due to their mass, heavy ions require low amounts of radiation to produce this damage and thus constitute the main radiation hazard for manned missions in space (\cite{Letaw1987, Hellweg2007}).

On the other hand, PUI relevance on the dynamics of the solar wind and coronal heating was theoretically analyzed in multiple references (\cite{Lee&Ip1987, Giacalone2000, Zank2016,  Isliker2017, Mostafavi2017, Zank2018}), showing that PUIs interact with the underlying low-frequency turbulence.
This generates Aflvénic fluctuations whose dissipation heats the solar wind plasma and may explain its non-adiabatic temperature profile.
Furthermore, multiple observations confirm the role played by PUIs in coronal extreme events (\cite{Russell2013, Lario2015}) and shocks (\cite{Kumar2018}).

In the inner heliosphere, PUIs have low density and are energetically unimportant compared to the much more dominant ambient particles.
It is then unlikely that they may develop mutually collective behaviour or be strongly scattered by self-generated wave-particle interactions.
These characteristics enable the use of test particle methods for their study (\cite{Luhmann_2003, Jarvinen2014}), where the convection electric field $\mathbf{E} = -\mathbf{V}_{sw}\times\mathbf{B}$ and the drift velocity $\mathbf{V}_D=\mathbf{E}\times\mathbf{B}/B^2$ play a central role (where $\mathbf{V}_{sw}$ is the solar wind velocity and $\mathbf{B}$ is the interplanetary magnetic field).

Electromagnetic fields generated in the turbulent solar wind are often called upon to explain the suprathermal behaviour of charged particles (\cite{Fermi1949, Matthaeus1984, Lazarin2012}).
Some theoretical treatments rely on the assumption of weak-turbulence (see \cite{Chandran2008} and references therein), where each field quantity is decomposed as the sum of a uniform background value plus a small-amplitude fluctuation.
This allows the turbulence to be described as a collection of waves, which may interact through triads or not at all.
Within this wave description, quasilinear theory (QLT) yields resonant conditions for the interaction between particles and the waves (\cite{Stix1992}), which may be used to obtain the evolution of particle density in phase space (\cite{Isenberg1983, Yoon1992, Marsch2001, Yoon2003, Isenberg2007, Moya2014}). Difficulties with the models based on this type of wave particle interaction are clearly exposed in \cite{Isenberg1983}, the main problem being the energy 
requirement at the high frequencies needed for the resonances which the weak turbulent cascade can not provide (see however alternatives based on a kinetic approach with different species populations as in \cite{Moya2014}).

A more realistic way to model the turbulence, not relying on a wave approximation, is through direct numerical simulations (DNS) of the magnetohydrodynamic (MHD) equations (\cite{Dmitruk_2004, Dmitruk2006a, LEHE2009, Dalena2014, Teaca2014, Gonzalez2016}), allowing for the self-consistent formation of coherent structures that occurs in a strong turbulence environment such as the solar wind.
However, most of these simulations only were evolved until a turbulent regime was reached and then a snapshot of the fields was taken to compute particle dynamics, ignoring the evolution of coherent structures and waves.
Although in the solar wind we expect the energy spectrum to reach a steady state, the fields still evolve in time and this affects particle dynamics.
While this may not be relevant when considering particles with high charge-to-mass ratio (i.e. electrons) whose characteristic times are much smaller than those of the turbulent fields, as charge-to-mass ratio decreases, we would not expect this to hold, as was shown for protons (\cite{Gonz_lez_2017}) and therefore dynamical evolution of the fields is desirable for the simulation of heavy ions. 

An alternative to the MHD approach for the fields has been recently shown in \cite{Kumar2017} where PIC (particle-in-cell) simulations are employed for the plasma (protons and electrons) with a population of heavy ions as test particles. Besides the usual limitation of PIC codes in terms of the realistic mass ratio of proton to electron, that study (\cite{Kumar2017}) is limited in the range of charge-to-mass ratio of the considered ions.

The main objective of the present paper is to study the effect of charge-to-mass ratio on test particle energization and displacement, particularly for heavy ions (lower charge-to-mass ratio),
covering a broad range of values which represent the type of heavy ions reported in observational studies \cite{Tracy2015}.
To this end, we perform a DNS of MHD turbulence and used their (dynamic) electromagnetic fields to simultaneously evolve charged particle populations with different charge-to-mass ratios.
We also apply a guide center model to understand particle displacement and their dependence (or lack thereof) with their charge-to-mass ratio.
We then perform a Voronoi tessellation (\cite{Monchaux2010, Obligado2014}) in order to calculate particle density and find regions of preferential concentration.
Combining this analysis with the guide center model, we identify structures able to trap and perpendicularly energize particles with an efficiency determined by their charge-to-mass ratio.

The paper is organized as follows:
In Section \ref{sec:model} we describe the model, equations and parameters used for the MHD fields and the test particles. The next Section \ref{sec:results} contains the results of the numerical simulations. 
In subsection \ref{ssec:disp_and_energ} we show particle energization and displacement for the different particle species.
In subsection \ref{ssec:GCM} we propose the guide center model, assess its validity and study the properties of the drift velocity and particle kinematics.
In subsection \ref{ssec:voronoi} we perform the Voronoi tessellation, studying cell volume distribution and investigating the underlying field responsible for clustering and its effect on energization.
Finally, we discuss our results and conclusions in Section \ref{sec:disc}.

\section{Models}{\label{sec:model}}

\subsection{The MHD fields}

The macroscopic description of the underlying plasma is given by compressible three-dimensional MHD: the continuity equation, the equation of motion and the magnetic field induction equation, supplemented by an equation of state for closure.
These are Eqs. (\ref{eq:mhd_continuity}-\ref{eq:mhd_state}) respectively and involve fluctuations of the velocity field $\textbf{u}$, magnetic field $\textbf{b}$ and density $\rho$.
We assume a background magnetic field $\mathbf{B}_0 = B_0\hat{z}$ such that $\mathbf{B}=\mathbf{B}_0+\mathbf{b}$

\begin{eqnarray}
\frac{\partial \rho}{\partial t} + \nabla\cdot(\rho \mathbf{u}) &=& 0 \label{eq:mhd_continuity}\\
\rho\left[ \frac{\partial \mathbf{u}}{\partial t} + (\mathbf{u}\cdot\nabla) \mathbf{u}\right] &=& -\nabla p + \frac{\mathbf{J}\times\mathbf{B}}{4\pi} + \mu\left( \nabla^2\mathbf{u} + \frac{1}{3}\nabla\left( \nabla\cdot\mathbf{u} \right) \right) \label{eq:mhd_motion} \\
\frac{\partial \mathbf{B}}{\partial t} &=& \nabla\times(\mathbf{u}\times\mathbf{B}) + \eta\nabla^2\mathbf{B} \label{eq:mhd_mag_induction} \\
\frac{p}{p_0} &=& \left(\frac{\rho}{\rho_0}\right)^\gamma \label{eq:mhd_state}
\end{eqnarray}

Here $p$ is the pressure, $\mathbf{J}=\nabla\times\mathbf{B}$ the current density, $\mu$ the viscosity and $\eta$ the magnetic diffusivity. 
In (\ref{eq:mhd_state}) we assume an adiabatic equation of state ($\gamma=5/3$), where $p_0$ and $\rho_0$ are respectively the reference pressure and density.
Although the Hall current is not taken into account in Eq. (\ref{eq:mhd_mag_induction}), it will be included later in the generalized Ohm’s law for the electric field and therefore impact particle motion equations.
This is because the dynamics of the fields described by Eqs. (\ref{eq:mhd_continuity}-\ref{eq:mhd_state}) are not notably affected by the Hall term, provided that the Hall scale is comparable to the dissipation scale (see \cite{Dmitruk2006b}).

The magnetic and velocity fields are expressed in Alfvén speed units based on the magnetic field fluctuations, defined as $v_0 = \delta B/\sqrt{4\pi\rho_0}$ with $\delta B=\langle b^2\rangle^{1/2}$.
The sonic Mach number $M = v_0/C_s $ defines the sound speed $C_s=\sqrt{\gamma p_0/\rho_0}$ in terms of the characteristic plasma speed.
As we are interested in a weakly compressible plasma, we choose $M=0.25$ for this simulation.
As a characteristic length, we use the isotropic MHD turbulence energy containing scale (or correlation length) $L= 2\pi \int (E(k)/k)dk / \int E(k)dk$ (where $E(k)$ is the energy spectral density at wavenumber $k$).
As the unit time scale, we use the eddy turnover time derived from the energy containing scale and the fluctuation Alfvén speed $t_0 = L/v_0$.

The MHD equations are solved numerically using a Fourier pseudospectral method with periodic boundary conditions in a cube of size $L_{box}=2\pi$; this scheme ensures exact energy conservation for the continuous time spatially discrete equations (\cite{MININNI2011316}).
The discrete time integration is done with a second-order Runge–Kutta method, and a spatial resolution of $N^3=512^3$ Fourier modes is used.
Aliasing is removed by the two-thirds rule truncation method, such that the maximum wavenumber resolved is $\kappa = N/3$.
This allowed us to reach values of kinematic Reynolds number $R=v_0 L\rho_0/\mu$ and magnetic Reynolds number $R_m=v_0 L/\eta$ of $R=R_m=2370$, in order to ensure resolution of the smallest scales ($\kappa > k_d $).
Here $k_d$ is the Kolmogorov dissipation wavenumber $k_d=(\epsilon_d/\mu^3)^{1/4}$ and $\epsilon_d$ the energy dissipation rate; defining a scale $l_d=2\pi/k_d$.

The system started from null velocity and magnetic perturbation fields ($\mathbf{u} = \mathbf{b} = 0$), constant density $(\rho=\rho_0)$ and $B_0=\delta B$.
In order to reach a turbulent steady state, the system is forced using external mechanical and electromotive forces $\mathbf{f}$ and $\nabla\times\mathbf{m}$ in Eqs. (\ref{eq:mhd_motion}) and (\ref{eq:mhd_mag_induction}), respectively.
This forcing is generated with random phases in the Fourier $k$-shells $2\leq k \leq 3$ every correlation time $\tau \approx t_0/18$.
Intermediate forcings, say at time $t$, are obtained through linear interpolation between the previous at $\lfloor t/\tau \rfloor\tau$ and the next at $\lceil t/\tau \rceil\tau$ (here, $\lfloor \bullet \rfloor$ and $\lceil \bullet \rceil$ denote the floor and ceil, respectively).
When a stationary turbulent state was reached, the background magnetic field was increased along with the electromotive forcing until the final value of $B_0/\delta B=9$.
In this stationary state, $L_{box}/L\approx 2.55$, $\langle u^2 \rangle\approx 2.8v_0^2$ and $l_d\approx L/60$.

\begin{figure}[ht]
\centering
\includegraphics[width=0.66\columnwidth]{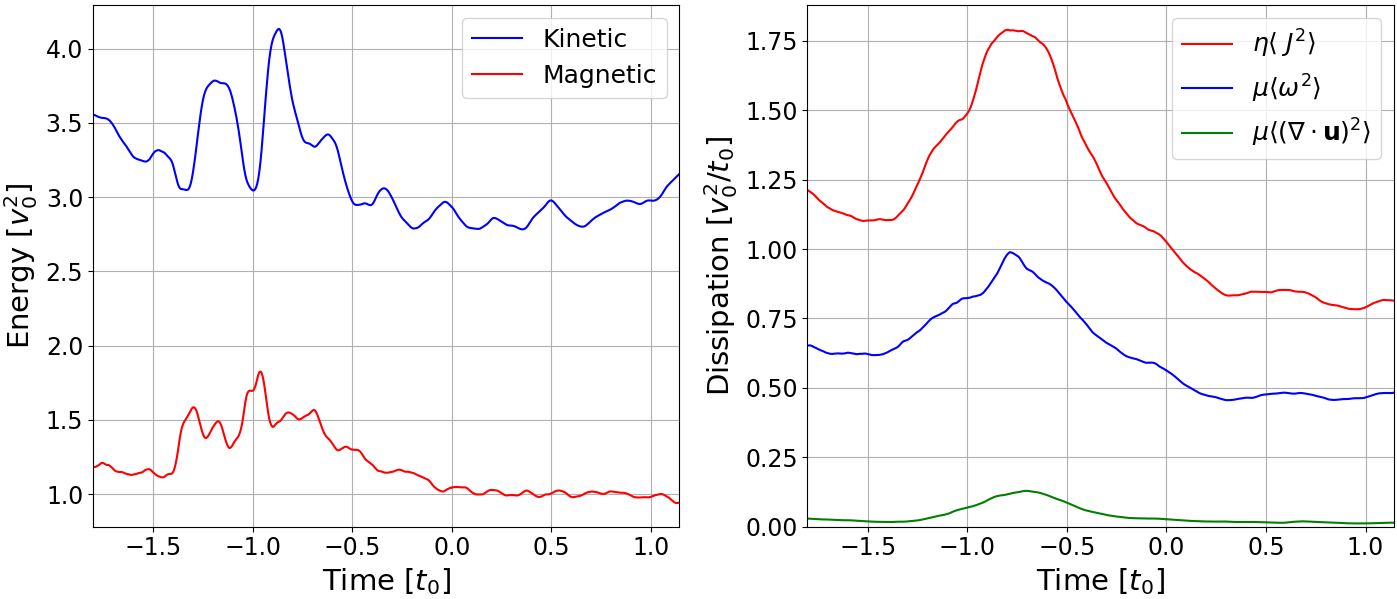}
\includegraphics[width=0.33\columnwidth]{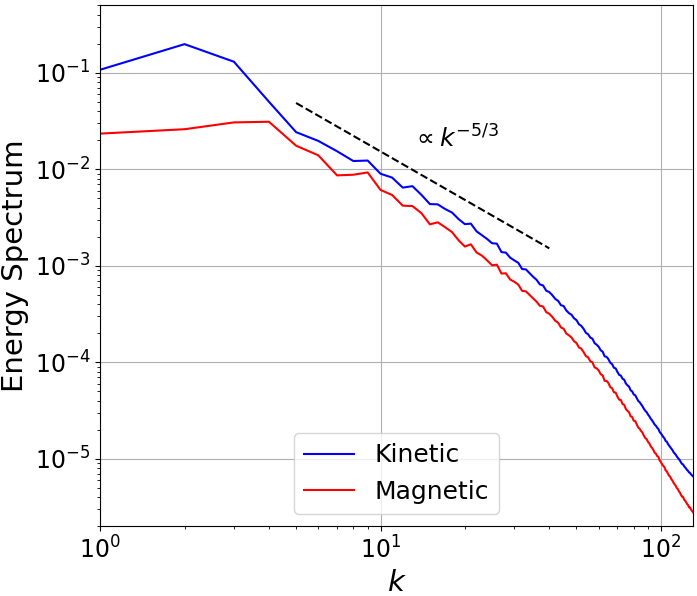}
\caption{Kinetic and magnetic energy (left) and dissipation terms (center) as a function of time, and energy spectrum during the stationary phase $t\geq 0$ (right).}
\label{fig:mhd_stats}
\end{figure}

\begin{figure}[ht]
\centering
\includegraphics[width=0.55\columnwidth]{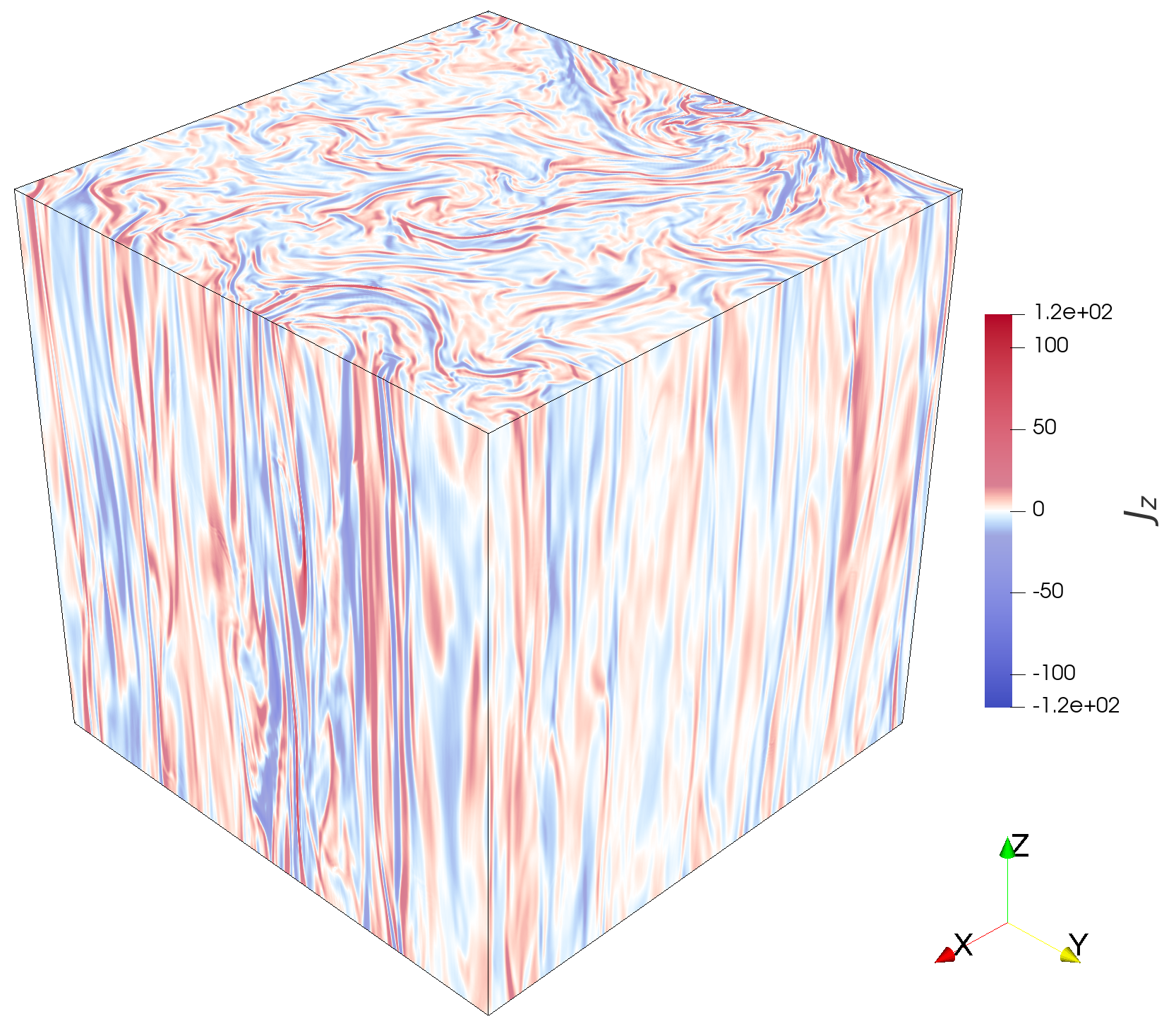}
\caption{Current density $J_z$ 3D view at $t=0.5t_0$, when turbulence is fully developed.}
\label{fig:current_sheets}
\end{figure}

The evolution of kinetic and magnetic energy is shown in Figure \ref{fig:mhd_stats}, along with their corresponding dissipations.
We choose $t=0$ as the time where the test particles (see below) are inserted into the system, right after the peak dissipation was achieved and both energy and dissipation reached a stationary state.
A mean energy spectrum during this period ($t\geq 0$) is also shown, with fluctuations being mostly negligible.
The final turbulent state can be seen in Figure \ref{fig:current_sheets}, where a 3D view of $J_z$ at time $t=0.5t_0$ is shown.
Current sheets are present with a characteristic thickness $\sim l_d$

\subsection{The particles}

In this stationary turbulent state, we release $5\times 10^5$ charged particles to evolve together with the dynamical MHD fields, which are unaffected by the particles (i.e., test particles).
Particles were uniformly distributed in the box and initialized with a Gaussian velocity distribution function, with a root mean square (rms) value between the Alfvén velocity and the rms of the velocity field $\langle v_i^2 \rangle^{1/2}\approx 1.5v_0$.
The dynamics of a test particle in an electromagnetic field are given by the nonrelativistic equation of motion:

\begin{equation}{\label{eq:newton}}
\frac{d\mathbf{r}}{dt} = \mathbf{v}, \quad \frac{d\mathbf{v}}{dt} = \alpha \left( \mathbf{E} + \mathbf{v}\times\mathbf{B} \right)
\end{equation}

The electric field $\mathbf{E}$ is obtained from the generalized Ohm's Law and its dimensionless (using a characteristic electric field $E_0=v_0B_0/c$) expression is

\begin{equation}{\label{eq:gen_ohm}}
\mathbf{E} = \frac{\mathbf{J}}{R_m} - \mathbf{u}\times\mathbf{B} + \frac{\epsilon}{\rho}\left( \mathbf{J}\times\mathbf{B} - \frac{1}{\gamma M^2}\nabla p_e \right)
\end{equation}

Here $\epsilon=\rho_{ii}/L$ is the Hall scale and relates the proton inertial length $\rho_{ii} = m_pc/\sqrt{4\pi\rho_0e^2}$ to the energy containing scale $L$, where $m_p$ and $e$ are the proton mass and charge, respectively.
In this work, we will identify the proton inertial length with the dissipation scale $\rho_{ii}=l_d$, given the solar wind observations supporting $\rho_{ii}\sim l_d$ (\cite{Leamon1998}).
For the electronic pressure $p_e$ we assume the plasma electrons and ions to be in thermal equilibrium $p_e=p_i=p/2$, with $p=p_e+p_i$.
While these last two terms give negligible contributions to the magnetic induction equation (\ref{eq:mhd_mag_induction}) through the curl of $\mathbf{E}$, they can be important at small scales comparable to the proton gyroradius (\cite{Dmitruk2006a}) and thus affect the dynamics of particles with similar gyroradii (see below).

The dimensionless parameter $\alpha$ in Eq. (\ref{eq:newton}) relates test particle and MHD field parameters through:
\begin{equation}{\label{eq:def_alpha}}
\alpha = Z\frac{m_p}{m}\frac{L}{\rho_{ii}}
\end{equation}
where $m$ and $Ze$ are the particles mass and charge, respectively.
This $\alpha$ is related to the charge-to-mass ratio and represents the gyrofrequency $\omega_g$ in a magnetic field of intensity $\delta B$.
Its inverse $1/\alpha$ represents the nominal gyroradius $r_g$ (in units of $L$) for particles with velocity $v_0$ in a magnetic field $\delta B$, and gives a measure of the range of scales involved in the system (from the outer scale of turbulence to the particle gyroradius).
Given the previous definitions, particles with $\alpha = L/\rho_{ii}$ will be considered ``protons'' and particles with lower $\alpha$ will be associated with ``heavy ions''.
These particles will be the focus of this work; those with nominal gyroradius above the dissipation scale $1/\alpha \geq l_d/L$, where $\epsilon=l_d/L\approx 1/60$.

The numerical integration of Eq. (\ref{eq:newton}) was done by a fourth-order Runge-Kutta method with adaptive time step. 
The values of the fields at each particle position are obtained by cubic splines in space from the three-dimensional grid of the MHD simulation.
The particle trajectories were integrated until the rms displacement is about $L_{box}/2$ (see below), to reduce the effects of the box periodicity on the motion.
The MHD fields evolved along with the particles but independently of them, such that the same fields were present for the different values of $\alpha$.

The chosen values for the different parameters are consistent with observations of the solar wind.
According to \cite{Andres2021}, typical values of $B_0/\delta B$ range from $2$ to $10$ with $\delta B \sim 2$nT; for $\rho_0 = m_p n \sim 3\times 10^{-26}$kg/m$^3$ we obtain $v_0\sim 10$km/s, which is consistent with $|\mathbf{u}|\sim v_0$.
In \cite{Alexandrova2009} a value of $l_d \approx \rho_{ii}\sim 70$km, in our case implying an energy scale $L\sim 400$km, a domain size of $L_{box}\sim 1000$km and a turnover time of $t_0\sim 40$s.

\section{Results}{\label{sec:results}}

\subsection{Displacement and energization}{\label{ssec:disp_and_energ}}

\begin{figure}[ht]
\centering
\includegraphics[width=\columnwidth]{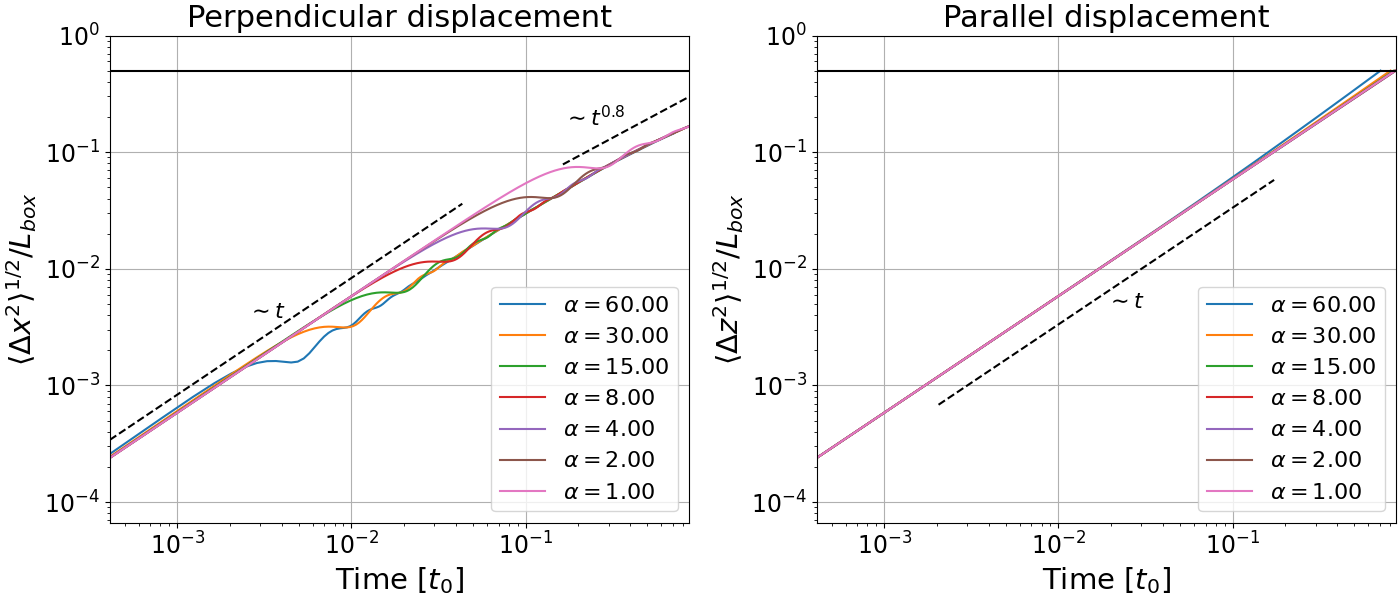}
\caption{Rms displacement of test particles in the perpendicular direction (left) and parallel direction (right) for all values of $\alpha$.}
\label{fig:disps_vs_time}
\end{figure}

In Figure \ref{fig:disps_vs_time} the root mean squared (rms) displacement in a transverse direction $x$ (there is no difference when $y$ is chosen instead) and the parallel direction $z$ is shown. 
In the left panel, the perpendicular displacement starts in a ballistic fashion, later collapsing to a sub-ballistic curve at different times.
While both regimes are identical for each $\alpha$, the switching times are proportional to the gyroperiod $\tau_\alpha\sim 1/\alpha$, suggesting that this initial behaviour is due to gyromotion effects.
During the latter sub-ballistic displacement, oscillations disappear and the curves become indistinguishable.
This suggests, as we will show later, that the mechanism of perpendicular diffusion is mostly independent of $\alpha$ as time advances.
This is not the case for the parallel displacement on the right panel, where the slope of the curves increases slightly with $\alpha$.
Furthermore, the parallel displacements display a ballistic or super-ballistic diffusion through the whole simulation.
All curves in Figure \ref{fig:disps_vs_time} are drawn until $\langle \Delta z^2 \rangle^{1/2} \approx L_{box}/2$, when the simulation is stopped. 
We selected this criteria to define the exit time $t_e^\alpha$ because the parallel component of the displacement is dominant ($\langle \Delta z^2 \rangle^{1/2} \sim 3 \langle \Delta x^2 \rangle^{1/2}$).
This duration was enough for particles to interact with multiple structures and results are not qualitatively affected when $\langle \Delta z^2 \rangle^{1/2} \approx L_{box}$ is chosen instead.

\begin{figure}[ht]
\centering
\includegraphics[width=\columnwidth]{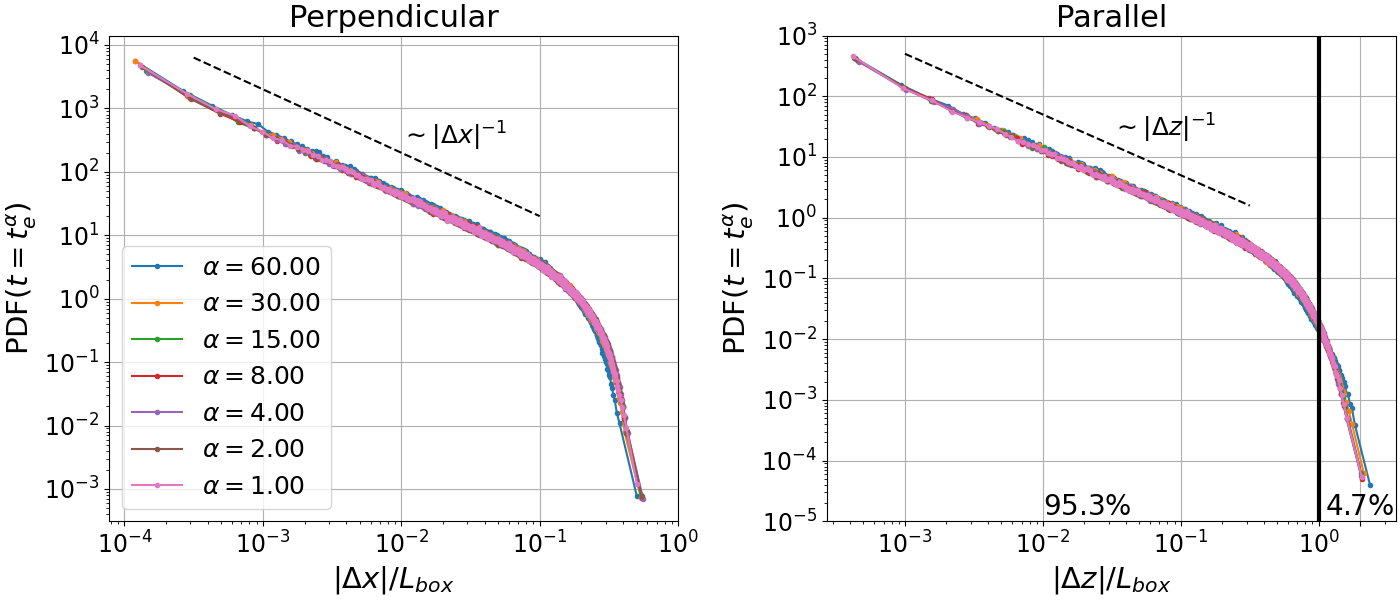}
\caption{Probability density functions of rms displacements at exit times $t=t_e^{\alpha}$ for perpendicular (left) and parallel (right) direction. Vertical line marks the box size.}
\label{fig:hist_disps_final}
\end{figure}

The exit times $t_e^{\alpha}$ are (slowly) decreasing with $\alpha$, ranging from $t_e^{(60)}\approx 0.71t_0$ to $t_e^{(1)}\approx 0.86t_0$.
At those times, no particle had traveled the whole box in the $x$-direction and less than $5\%$ of the particles had done so in the $z$-direction, as can be seen in the displacement probability density functions (PDFs) of Figures \ref{fig:hist_disps_final}.
For all $\alpha$, PDFs exhibit a power-law with exponent $-1$ at low displacements, giving way to an exponential decay for high displacements.

\begin{figure}[ht]
\centering
\includegraphics[width=\columnwidth]{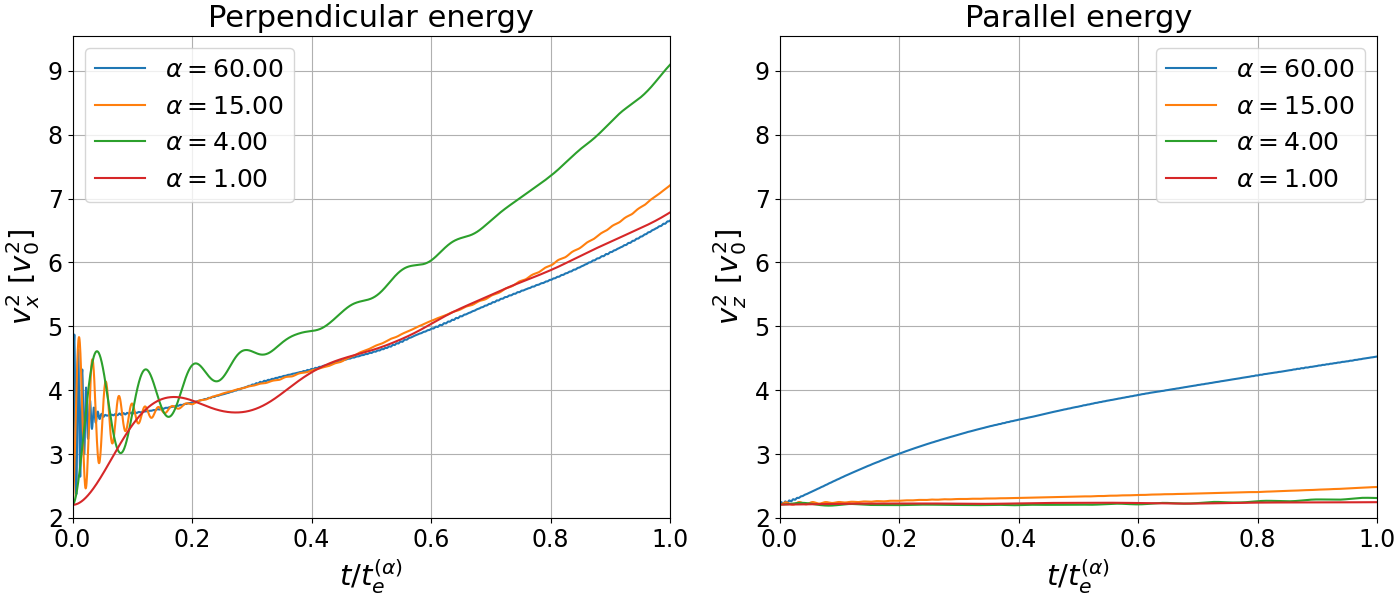}
\caption{Evolution of the mean kinetic energy for the perpendicular (left) and parallel (right) components, for some chosen values of $\alpha$. Time is normalized by the exit time $t_e^{\alpha}$.}
\label{fig:energy_vs_time}
\end{figure}

Now, we turn our attention to the kinetic energy of the particles, which is shown in Figure \ref{fig:energy_vs_time} for some values of $\alpha$.
The left panel corresponds to the perpendicular energy $v_x^2$ (as before, $v_y^2$ is identical) while the right panel corresponds to parallel energy $v_z^2$.
Although both components start with the same mean energy $\langle v_i^2 \rangle_0\approx 2.2v_0^2$, the perpendicular component has a quick oscillatory increase until the oscillations cease (as before, related to the gyromotion) around a value of the fluid kinetic energy $\langle u^2 \rangle$, followed by a steady increase.
In contrast, parallel energization is increasing with $\alpha$ but almost negligible except for the highest values of $\alpha$.
As expected for particles with $\alpha^{-1} \geq l_d/L$, the energization is dominant in the perpendicular component but the $\alpha$-dependence is not so clear anymore.

\begin{figure}[ht]
\centering
\includegraphics[width=\columnwidth]{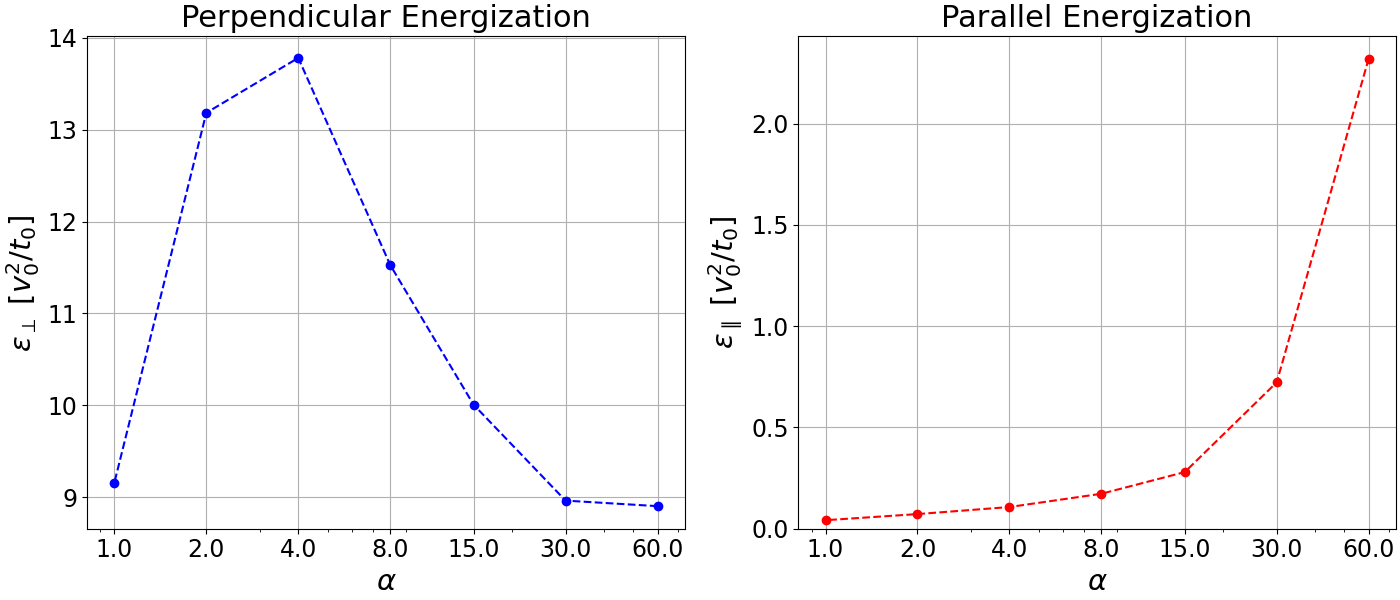}
\caption{Mean energization rates as a function of $\alpha$ for the perpendicular (left) and parallel (right) component.}
\label{fig:energy_rate_vs_alpha}
\end{figure}

In order to understand this issue, we plot the rate of perpendicular (left panel) and parallel (right panel) energization rates in Figure \ref{fig:energy_rate_vs_alpha}, defined as
\[ \varepsilon_\perp = \frac{\langle \Delta v_\perp^2 \rangle_e^{\alpha}}{t_e^{\alpha}} \quad,\quad \varepsilon_\parallel = \frac{\langle \Delta v_\parallel^2 \rangle_e^{\alpha}}{t_e^{\alpha}}\]
where $\langle \Delta v_j^2 \rangle_e^{\alpha}$ is the difference between the mean energy at time $t_e^{\alpha}$ and the initial mean energy.
In the perpendicular component there seems to be a maximum for $\alpha = 4$, while the parallel component is increasing with $\alpha$.

As previous works have shown (\cite{Dmitruk_2004, LEHE2009}), the energization tends to be mainly parallel for high values of $\alpha\sim\omega_g$ and mainly perpendicular for low
values of $\alpha$.
This is due to the interaction of the particles with the current sheets and other structures with size $\sim l_d$ (\cite{Gonzalez2016, Gonz_lez_2017}).
Low $\alpha$ particles have high gyroradius $r_g$ and are unable to exploit the coherence of $J_z$ inside a current sheet as do particles with $r_g\lesssim l_d$.
On the other hand, for $r_g \lesssim l_d$ the particles inside a current sheet experience almost constant $\mathbf{E}_\perp$, whose net force averages to $0$ in a given gyroperiod and amounts to a small perpendicular energization.
Furthermore, the increase in $r_g$ allows particles to sample bigger regions of the plasma in search of higher $\mathbf{E}_\perp$ gradients.
So, the perpendicular energization increases as $\alpha$ decreases.
Nonetheless, this cannot hold indefinitely, for we know that in the limit $\alpha \to 0$ there is no interaction and both energizations must be null.
As $\varepsilon_\perp(\alpha)$ is a non-zero smooth function with $\varepsilon_\perp=0$ for $\alpha=0$ and $\alpha\to\infty$, there must be at least one maximum $\alpha_c$.
With this in mind, we expect the perpendicular energization for very small $\alpha$ to increase as the intensity of the interaction increases up to this critical value $\alpha_c$.
From this $\alpha_c$ onward, the reduction in $r_g$ begins to nullify the effect of $\mathbf{E}_\perp$.

\begin{figure}[ht]
\centering
\includegraphics[width=0.55\columnwidth]{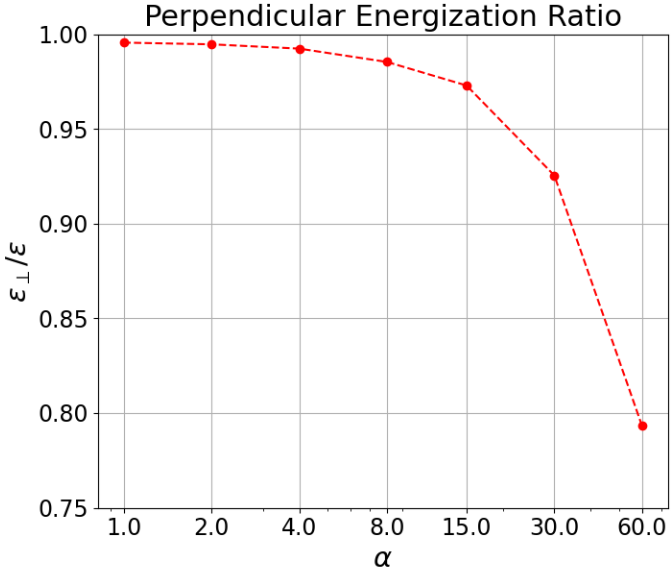}
\caption{Ratio between perpendicular and total energization as a function of $\alpha$.}
\label{fig:energy_ratios_vs_alpha}
\end{figure}

In summary, the competition between interaction strength (relevant at high $\alpha$) and exploration/exploitation (relevant at low $\alpha$) yields one or more maxima for $\varepsilon_\perp(\alpha)$. In the context of this work, it would seem one such maxima is $\alpha_c\approx 4$.
In order to free our analysis from this competition, we plot the ratio of perpendicular to total energization in Figure \ref{fig:energy_ratios_vs_alpha}; the fraction of energy that ends in the perpendicular component, irrespective of the total amount.
In this way, we remove the interaction strength factor and focus only at the geometric/structural aspect, which clearly benefits parallel energization as $\alpha$ increases.
It is in terms of this interpretation that we are able to assert that perpendicular energization becomes dominant (albeit not necessary higher) for low values of $\alpha$.

\subsection{A guide center model}{\label{ssec:GCM}}

As shown in Figure \ref{fig:disps_vs_time}, the behaviour of the perpendicular displacement becomes independent of $\alpha$ and raises the question of which mechanism is responsible for this.
In other words, we are searching for a mechanism which gives the particles a velocity independent of their $\alpha$ and the drift velocity in a constant electric and magnetic field springs to mind (\cite{RoedererZhang})

\begin{equation}{\label{eq:def_VD}}
    \mathbf{V}_D = \frac{\mathbf{E}\times \mathbf{B}}{B^2}
\end{equation}

\begin{figure}[ht]
\centering
\includegraphics[width=0.55\columnwidth]{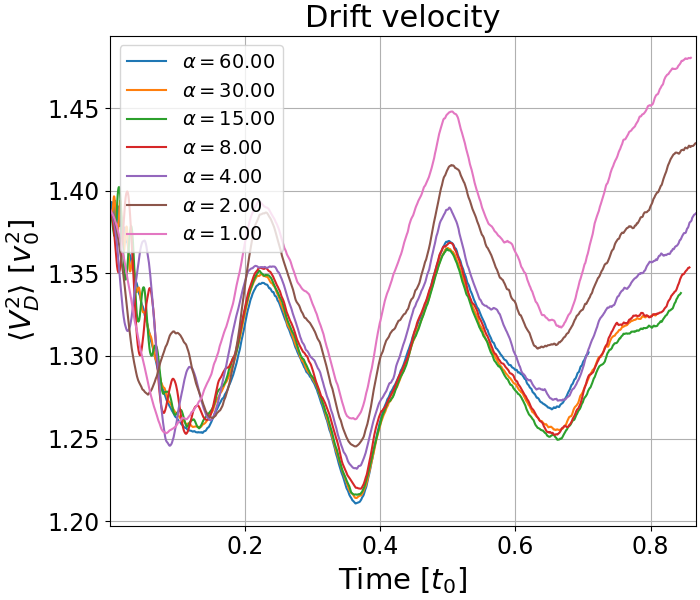}
\caption{Mean squared value of the drift velocity experienced by test particles as a function of time, for all values of $\alpha$.}
\label{fig:rms_VD_vs_time}
\end{figure}

While the magnetic field $\mathbf{B}$ could be approximately constant given the strong guide field $\mathbf{B}_0$, this is not immediate for the electric field $\mathbf{E}$, even for particles with $r_g \lesssim l_d$.
This approximation models particle motion as long as the characteristic times and scales of the fields are much greater than the gyroperiod and gyroradius of the particles, respectively.
Even though the dependence on $\alpha$ is not explicit in Eq. (\ref{eq:def_VD}), it may be implicit in the positions $\mathbf{x}_i$ of the particles, affecting the $\mathbf{V}_D$ experienced by the particle (interpreting $\mathbf{V}_D(\mathbf{x})$ as a new vector field).
To answer this question, in Figure \ref{fig:rms_VD_vs_time} is shown the mean squared value of $\mathbf{V}_D$ experienced by the particles of different $\alpha$, confirming a weak dependence on $\alpha$.

\begin{figure}[ht]
\centering
\includegraphics[width=\columnwidth]{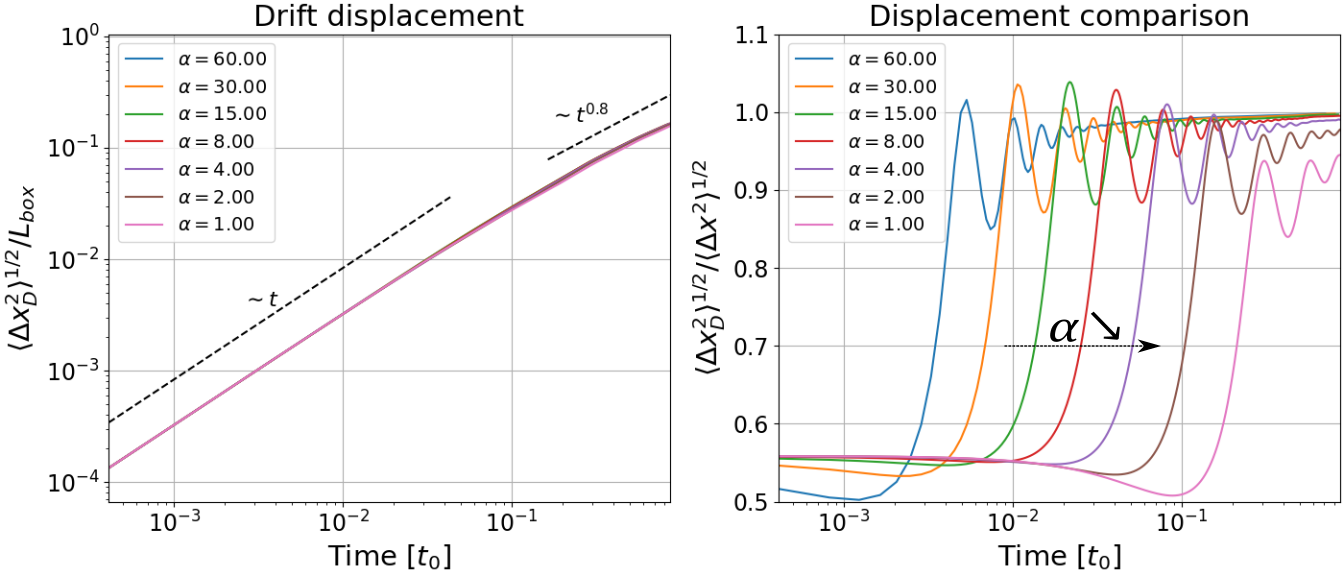}
\caption{As a function of time, (left) rms displacement predicted by the guide center model, and (right) its ratio to the actual rms displacement, for all values of $\alpha$.}
\label{fig:comp_rmsd_drift}
\end{figure}

In this approximation, $\mathbf{V}_D$ corresponds to the velocity of the guiding center of the particle. 
With this in mind, we propose to define a drift position $\mathbf{x}_{D}$ of the guiding center for each particle as 
\begin{equation}{\label{eq:def_xD}}
    \frac{d\mathbf{x}_{D, i}}{dt} = \mathbf{V}_D(\mathbf{x}_i, t) \qquad \mathbf{x}_{D, i} = \mathbf{x}_i(0) + \int_0^t\mathbf{V}_D(\mathbf{x}_i(t'), t') dt'
\end{equation}
where $\mathbf{x}_i(t)$ is the actual position of the particle.
Given that $\mathbf{x}_i(t)$ is a known function of time, we numerically integrated Eq. (\ref{eq:def_xD}) to obtain $\mathbf{x}_{D, i}(t)$.
The perpendicular displacement of this drift position is shown in the left panel of Figure \ref{fig:comp_rmsd_drift}, along with a plot of the ratio $\langle \Delta x_D^2 \rangle/\langle \Delta x^2 \rangle$ for comparison.
We see that this drift displacement lacks oscillations, which is consistent with its interpretation as the guide center.
Because of this fact, the discrepancy is most notorious for initial times, before the gyromotion is averaged out and the particle reaches its drift velocity; this explains the initial sudden increase in perpendicular energy in Figure \ref{fig:energy_vs_time}.
For latter times, the ratios for all $\alpha\geq 4$ converge within a $98\%$ to $1$, showing this guide center model to account for most of the displacement; even for $\alpha=1,2$ the discrepancy is less than $10\%$.
Its worth noting that the initial ballistic behaviour is captured by this model, showing that is not only due to the gyromotion.

\begin{figure}[ht]
\centering
\includegraphics[width=0.55\columnwidth]{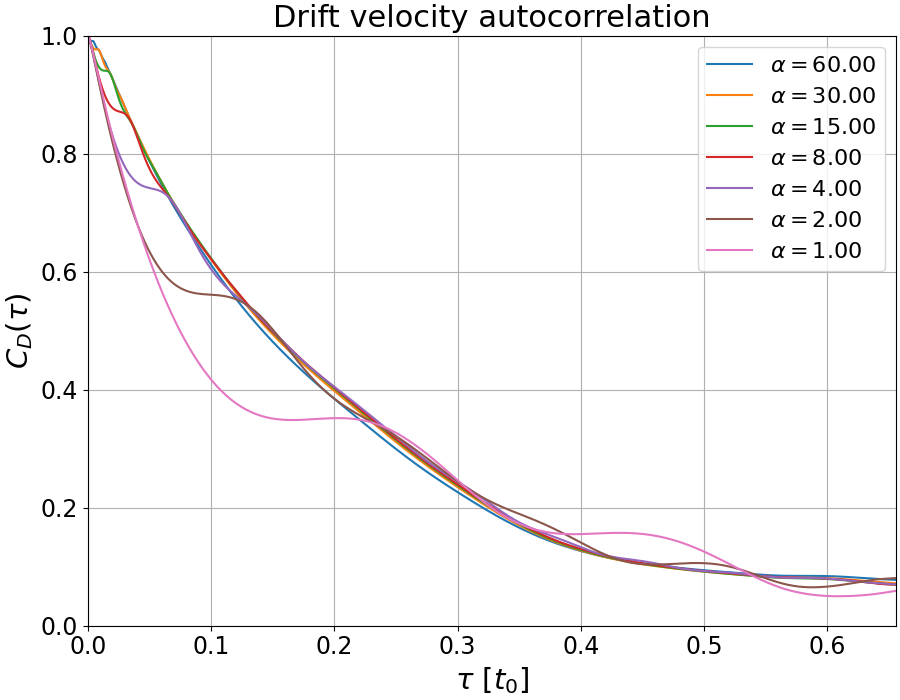}
\caption{Drift velocity autocorrelation as a function of time, for all values of $\alpha$. }
\label{fig:vd_corr_vs_time}
\end{figure}

Having shown the $\alpha$-independent drift as the dominant mechanism in perpendicular displacement, we now turn to the non trivial behaviour of the displacement; the initial ballistic and latter $\sim t^{0.8}$.
These super-diffusive exponents hint at some coherence in the (perpendicular) drift velocity in contrast to the uncorrelated velocities of Brownian motion.
To assert this, we calculate the drift velocity autocorrelation function
\begin{equation}
C_D(\tau) = \frac{\langle \mathbf{V}_{D\perp}(t)\cdot\mathbf{V}_{D\perp}(t+\tau) \rangle_t}{\langle |\mathbf{V}_{D\perp}(t)|^2\rangle_t}    
\end{equation}
where $\langle \bullet \rangle_t$ represents the average over time $0\leq t \leq t_e^{\alpha}-\tau$ and over all particle trajectories.
The resulting autocorrelation functions are shown in Figure \ref{fig:vd_corr_vs_time} and
seem to be weakly dependent on $\alpha$.
As before, there is an initial behaviour for $t\lesssim \tau_\alpha$ (the particle gyroperiod) due to particle gyromotion followed by a collapse to a common behaviour, most probably related to the evolution of $\mathbf{V}_D$.
Defining the time when $C_D(\tau_c)=0.1$ as the correlation time, we obtain a value comparable to the turn-over time $\tau_c\approx t_0/2$, reinforcing its interpretation as a characteristic time of the drift velocity field. 
For most values of $\alpha$, this correlation time is much greater than the particle gyroperiod $\tau_c / \tau_\alpha \approx 1.8\alpha$, which explains the increased coherence in the drift motion and provides a tool for quantifying the applicability of the guide center model.
The lower values of $\alpha=1,2$ have $\tau_\alpha \sim \tau_c$, which explains the discrepancies seen in Figure \ref{fig:comp_rmsd_drift}, in contrast to higher values of $\alpha$.

So far we have refrained from analysing parallel drift velocity and displacement because particle gyromotion is almost completely confined to the perpendicular plane.
As shown in Figure \ref{fig:disps_vs_time}, motion along the parallel direction is ballistic or slightly super-ballistic, which cannot be explained by a stationary drift velocity.
To show this, we calculated the mean direction of each particle velocity (and for each component) throughout its evolution, using the sign function

\begin{equation}
    \langle \text{sg}(\mathbf{v}_j) \rangle = \frac{1}{t_e^\alpha}\int_0^{t_e^\alpha} \text{sg}(\mathbf{v}_j(t)) dt \approx \frac{1}{N_{steps}} \sum_{i=1}^{N_{steps}} \text{sg}(\mathbf{v}_j(t_i))
\end{equation}

\begin{figure}[ht]
\centering
\includegraphics[width=\columnwidth]{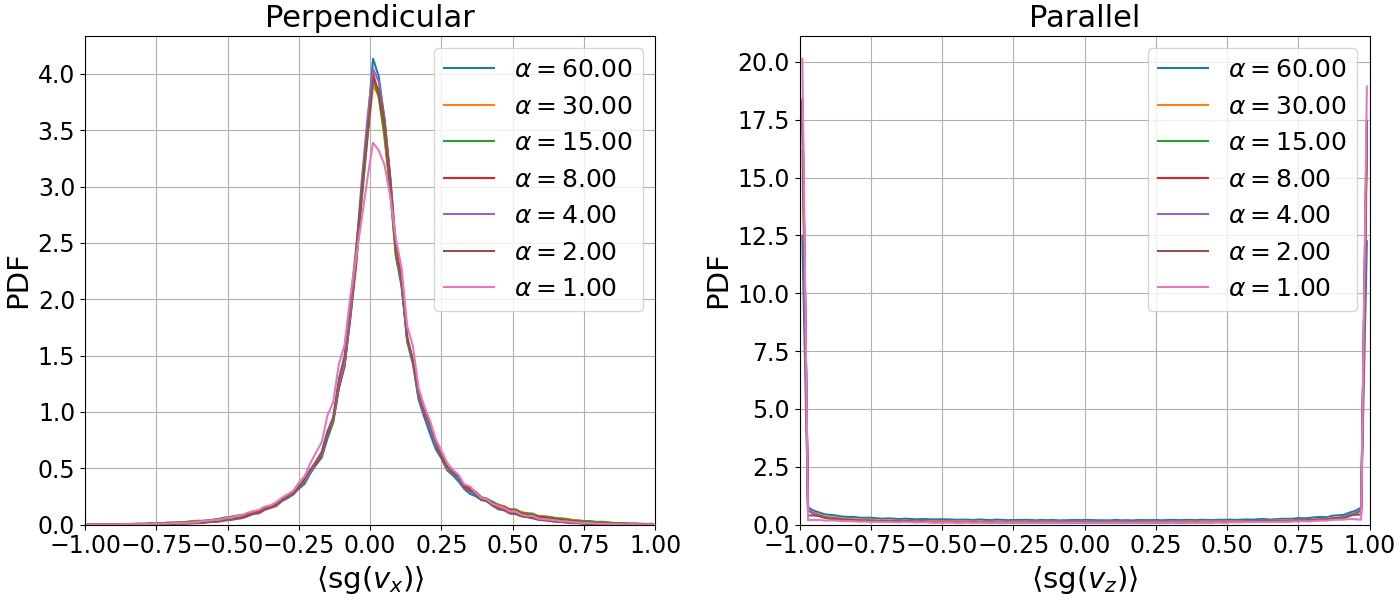}
\caption{Probability density functions of the mean velocity direction for (left) the perpendicular component and (right) the parallel component, for all values of $\alpha$.}
\label{fig:hist_signs}
\end{figure}

In Figure \ref{fig:hist_signs} are shown PDFs of $\langle \text{sg}(v_x) \rangle$ and $\langle \text{sg}(v_z) \rangle$, displaying wildly different behaviour.
The velocity direction in the perpendicular plane shifts quickly due to the gyration of particles, resulting in a steep peak in $\langle \text{sg}(v_x) \rangle=0$.
This is specially true as time pases, kinetic energy increases and drift velocity becomes negligible.
On the other hand, in the parallel direction the particles are concentrated around $\langle \text{sg}(v_z) \rangle= \pm 1$, showing that they mostly adhere to their original direction.
The non-zero distribution outside $\langle \text{sg}(v_z) \rangle \approx \pm 1$ is evidence of the existence of acceleration not only increasing parallel energy but also changing movement direction.

\begin{figure}[ht]
\centering
\includegraphics[width=\columnwidth]{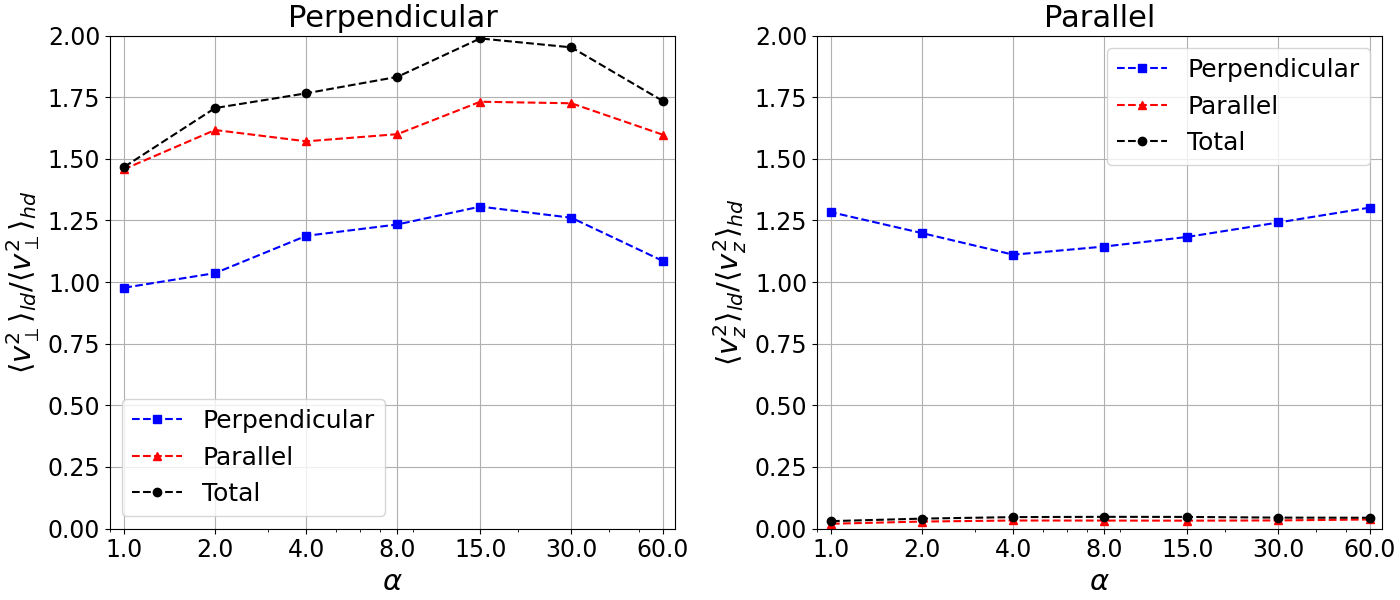}
\caption{As a function of $\alpha$, ratio of the final mean energy between low displacement (ld) and high displacement (hd) particles, both in the perpendicular (left) and parallel (right) components.}
\label{fig:energy_hd_ld}
\end{figure}

To conclude this section, we try to find some relation between displacement and energization. At final times $t_e^{\alpha}$, we calculated the displacement and energy (in each component) for all particles and separated them according to the former.
We took the $20\%$ particles with higher displacements (hd) and the $20\%$ with lower displacements (ld) according to each direction and then combined (total displacement).
We then calculated their mean energy, whose ratio is shown in Figure \ref{fig:energy_hd_ld} for the different values of $\alpha$.
In the right panel, we see that parallel energization is increased by high parallel displacement but diminished by high perpendicular displacement.
The former result is consistent with Figure \ref{fig:hist_signs}, as particles that do not change movement direction have greater $|\Delta z|$ as $|v_z|$ increases.
The latter could be understood by considering that parallel energization is mainly due to current sheets, which are elongated in the parallel direction but thin in the perpendicular direction.
Therefore, perpendicular displacement can easily pull particles off these sheets, reducing their parallel energization.
On the other hand, perpendicular energization is increased by low displacement, mainly in the parallel direction.
We leave the interpretation of this fact for the next section.

\subsection{Preferential concentration}{\label{ssec:voronoi}}

In this section, we turn our attention to the slight discrepancies in the drift velocity experienced by different values of $\alpha$.
As said, given that all particles for all $\alpha$ are immersed in the exact same fields and that the drift velocity defined in Eq. (\ref{eq:def_VD}) is independent of $\alpha$, the only way particles with different $\alpha$ could experience different drift velocities is through their spatial distribution.
If particles retained their initial uniformly distribution in space, all curves in Figure \ref{fig:rms_VD_vs_time} should collapse.
Thus, we should expect a deviation from uniformity, which we could identify with regions of preferential concentration.

\subsubsection{Voronoi tesselation and clustering}

To quantify the preferential concentration, we employ the Voronoi tessellation method and compare volume statistics against a Random Poisson Process (RPP) (\cite{Monchaux2010, Obligado2014, UHLMANN2020, Reartes2021}).
We performed this analysis at certain times, taking all particle positions $\mathbf{x}_i$ and performing a Voronoi tessellation.
This process provides a cell for each particle, whose volume $\mathcal{V}_i$ can be interpreted as the inverse of the local particle density.
In the context of this work, local particle density is associated with the probability (density) of finding a test particle in the region.
Therefore, high particle volumes are associated to voids and low particle volumes to clusters.

\begin{figure}[ht]
\centering
\includegraphics[width=\columnwidth]{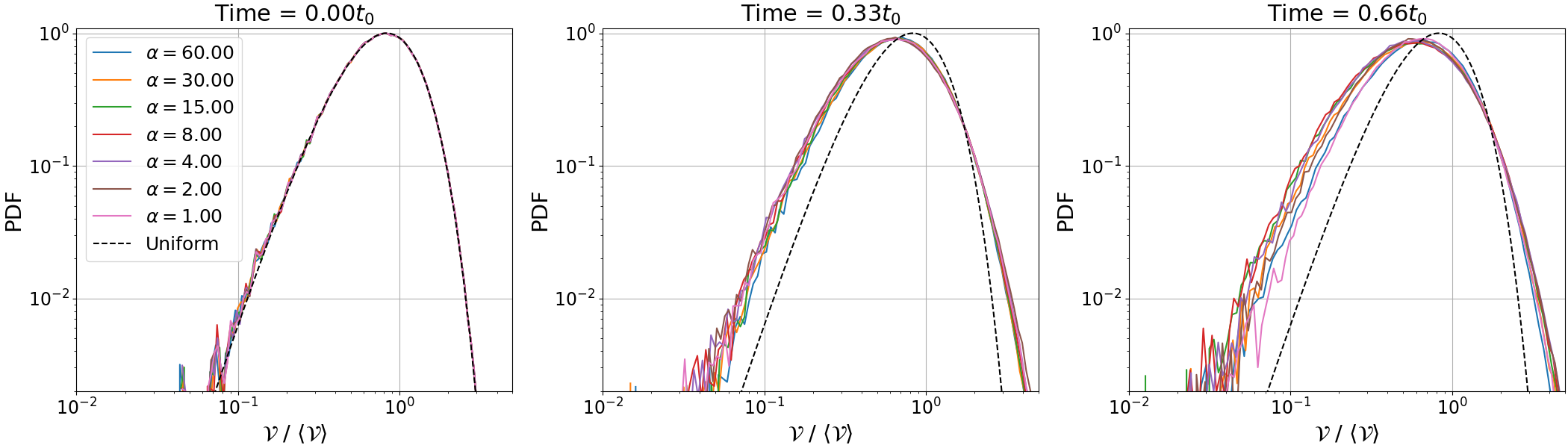}
\caption{Probability density function of (normalized) cell volume distribution at different times, for all values of $\alpha$. Dashed line corresponds to an uniform distribution.}
\label{fig:hists_volumes}
\end{figure}

We then calculated the PDF of the normalized volumes $\mathcal{V}/\langle \mathcal{V} \rangle$ and compared it to the PDF of a RPP, which corresponds to the uniform case.
In Figure \ref{fig:hists_volumes} we show these PDFs for all $\alpha$ at times $t=0$ (initial), $t=0.33t_0$ (intermediate) and $t=0.66t_0$ (close to final).
As time progresses, the distribution deviates from that corresponding to the uniform case, although not at the same rate for all $\alpha$.

\begin{figure}[ht]
\centering
\includegraphics[width=0.55\columnwidth]{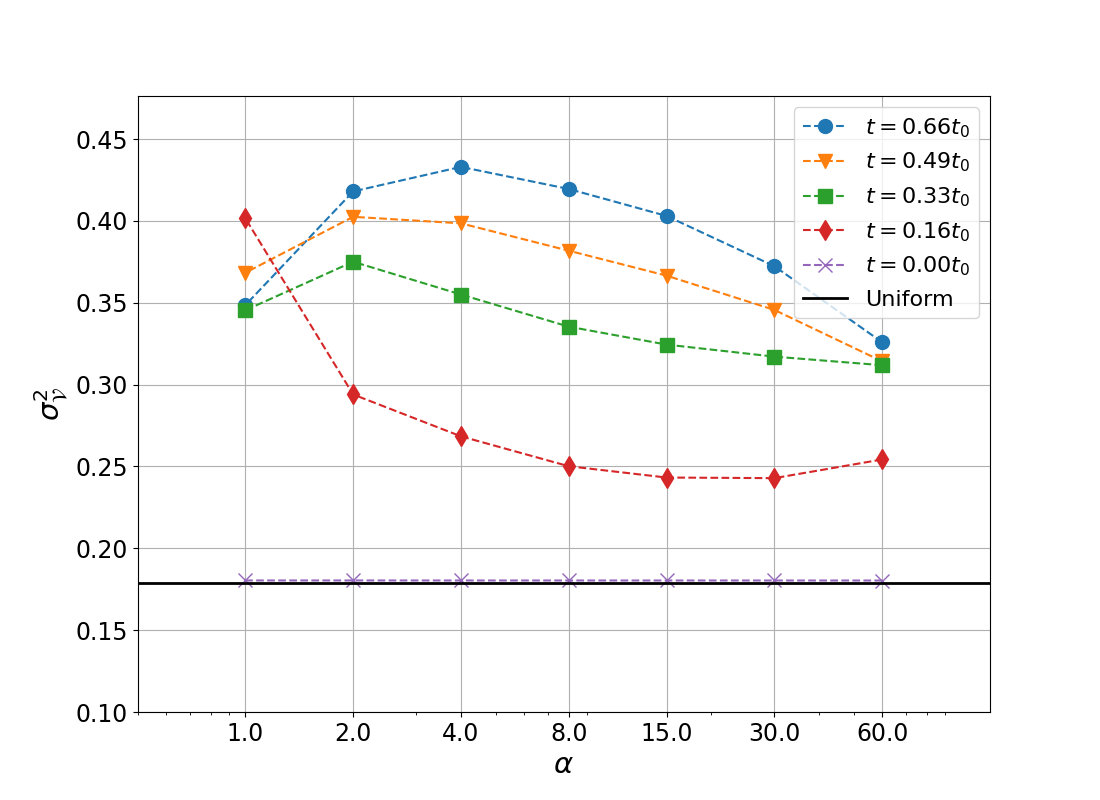}
\caption{Standard deviation of normalized cell volume distribution as a function of $\alpha$, at different times. Dashed line marks the value corresponding to a uniform distribution.}
\label{fig:sigma_volumes}
\end{figure}

In order to quantify this notion, we computed the normalized volume variance $\sigma_\mathcal{V}^2$ for each distribution and compared them to the known value for a RPP $\sigma_\mathcal{V}^2\approx 0.18$.
This variance $\sigma_\mathcal{V}^2$ should increase as the PDFs stray from that of a uniform distribution, as we see in Figure \ref{fig:sigma_volumes}.
For most values of $\alpha$, $\sigma_\mathcal{V}^2$ increases with time at a decreasing rate, showing an explosive rearranging at the start followed by a slow settlement. 
At final time $t=0.66t_0$ the curve has a clear maximum, again located at $\alpha=4$, which induces to analyze the relation between energization and clustering.

To do so, we need a way of determining which particles are clustered and which are not.
Following (\cite{Monchaux2010, Obligado2014}), we consider a particle $i$ to be clustered if its volume $\mathcal{V}_i$ is lower than a certain critical value $\mathcal{V}_c$.
This $\mathcal{V}_c$ is chosen as the left-most intersection between the PDF of that $\alpha$ and the PDF of the uniform case, as seen in Figure \ref{fig:hists_volumes}.
Its not hard to see that for these slight deviations, the fraction of clustered particles by this criteria will be considerable ($\sim 20\%$).

\begin{figure}[ht]
\centering
\includegraphics[width=\columnwidth]{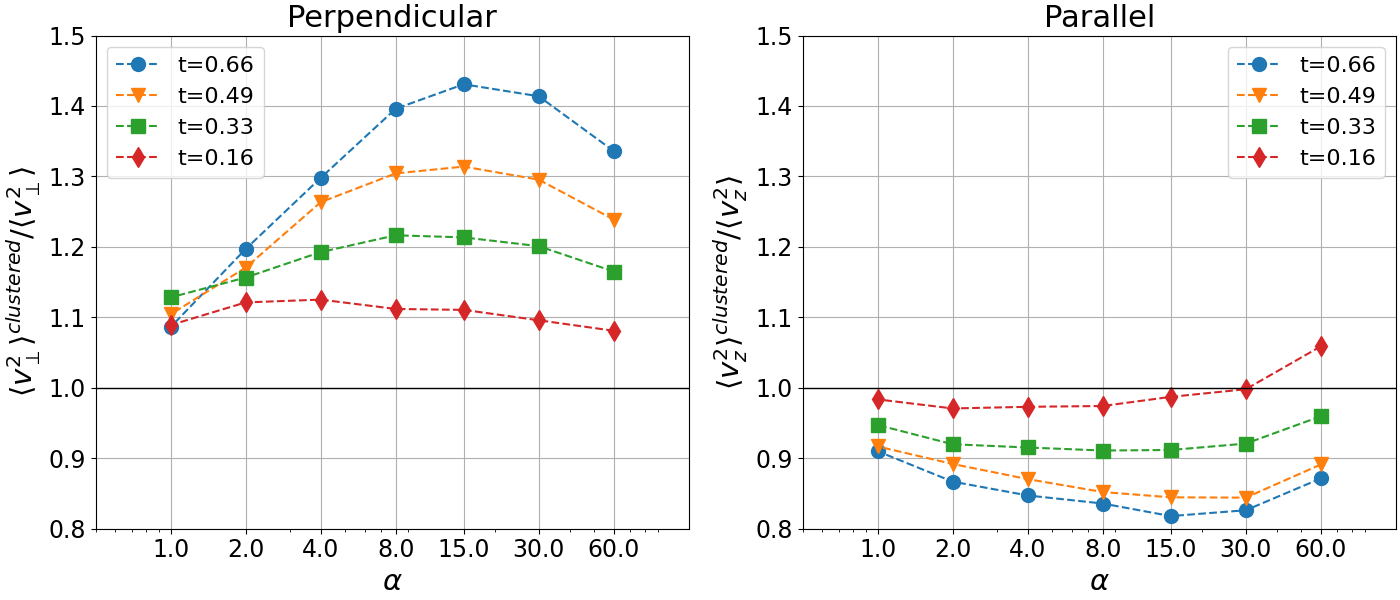}
\caption{Ratio of the mean energy between clustered particles and the whole population as a function of $\alpha$ at different times, for both perpendicular (left) and parallel (right) components.}
\label{fig:clustered_energ}
\end{figure}

We applied this criteria for all $\alpha$ and some chosen times, and calculated the mean energies $\langle  v_\perp^2 \rangle^{clustered}$ and $\langle  v_z^2 \rangle^{clustered}$ of the clustered particles only.
The ratio between this energies and the mean energies of the whole system are shown in Figure \ref{fig:clustered_energ}.
Clearly, clustered particles have (on average) more perpendicular energy and less parallel energy, even more so as time progresses.
This suggests the existence of regions where particles concentrate and benefit from a larger perpendicular energization.

So far we have identified 3 distinct phenomena in test particle dynamics:

\begin{itemize}
    \item[(a)] High (low) perpendicular (parallel) energization.
    \item[(b)] Low displacement (in either direction).
    \item[(c)] Preferential concentration (clustering).
\end{itemize}

We have related (a) with (b) through Figure \ref{fig:energy_hd_ld} and with (c) through Figure \ref{fig:clustered_energ}.
This suggests that (b) and (c) should be related too and we could expect clustering to arise from trapping, which would also reduce displacement.

\begin{figure}[ht]
\centering
\includegraphics[width=\columnwidth]{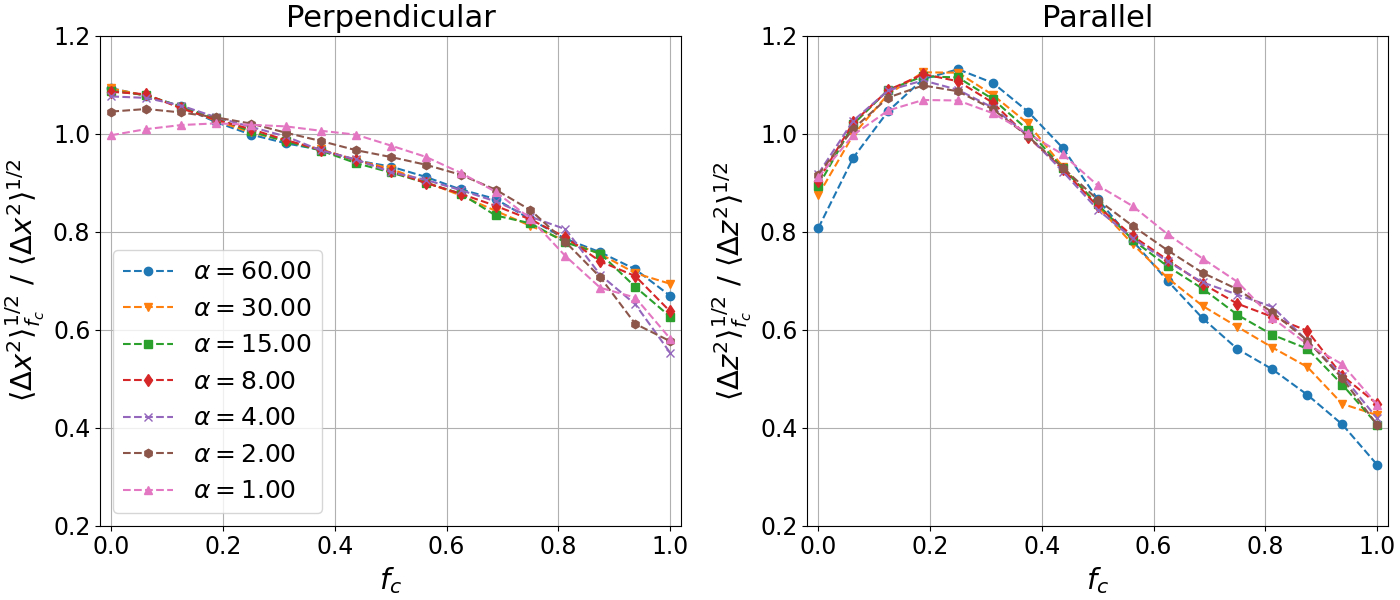}
\caption{Rms displacement (relative to that of the whole population) as a function of the fraction of clustered time for both the perpendicular (left) and parallel (right) components, for all values of $\alpha$.}
\label{fig:disp_vs_fc}
\end{figure}

To confirm this intuition, we calculate the fraction of time $f_c$ that each particle is clustered during the simulation.
We then separate all particles according to their $f_c$, calculated the final (at $t=t_e^\alpha$) displacement for each group and compare that of the whole population.
In Figure \ref{fig:disp_vs_fc} we can see a mostly decreasing relation between displacement and $f_c$, with very little dependence on $\alpha$.
This tells us that particles travel much less (up to $\sim 50\%$) as they spend more time clustered; they are trapped in these regions.
Nonetheless, the maximum around $f_c\approx 0.2$ in the parallel component shows that particles that are \textit{never} clustered also travel less ($\sim 15\%$).

In this section we have shown the existence of regions that trap and perpendicularly energize particles.
But, what is the mechanism behind this phenomenon?
What is special about these regions?

\subsubsection{Understanding the clustering criteria}

In order to answer the last questions, we investigate the fields experienced by clustered particles in contrast to those of a (hypothetical) uniform distribution in the box.
To understand the clustering, we could start by noting that Eq. (\ref{eq:def_xD}) is the equation of a tracer particle in a velocity field $\mathbf{V}_D(\mathbf{x}, t)$.
It is known that tracer particles tend to accumulate in (vacate) regions with negative (positive) divergence (\cite{Falkovich2001Rev,Balkovsky2001,Bec2004,Dhanagare2014}).

\begin{figure}[ht]
\centering
\includegraphics[width=\columnwidth]{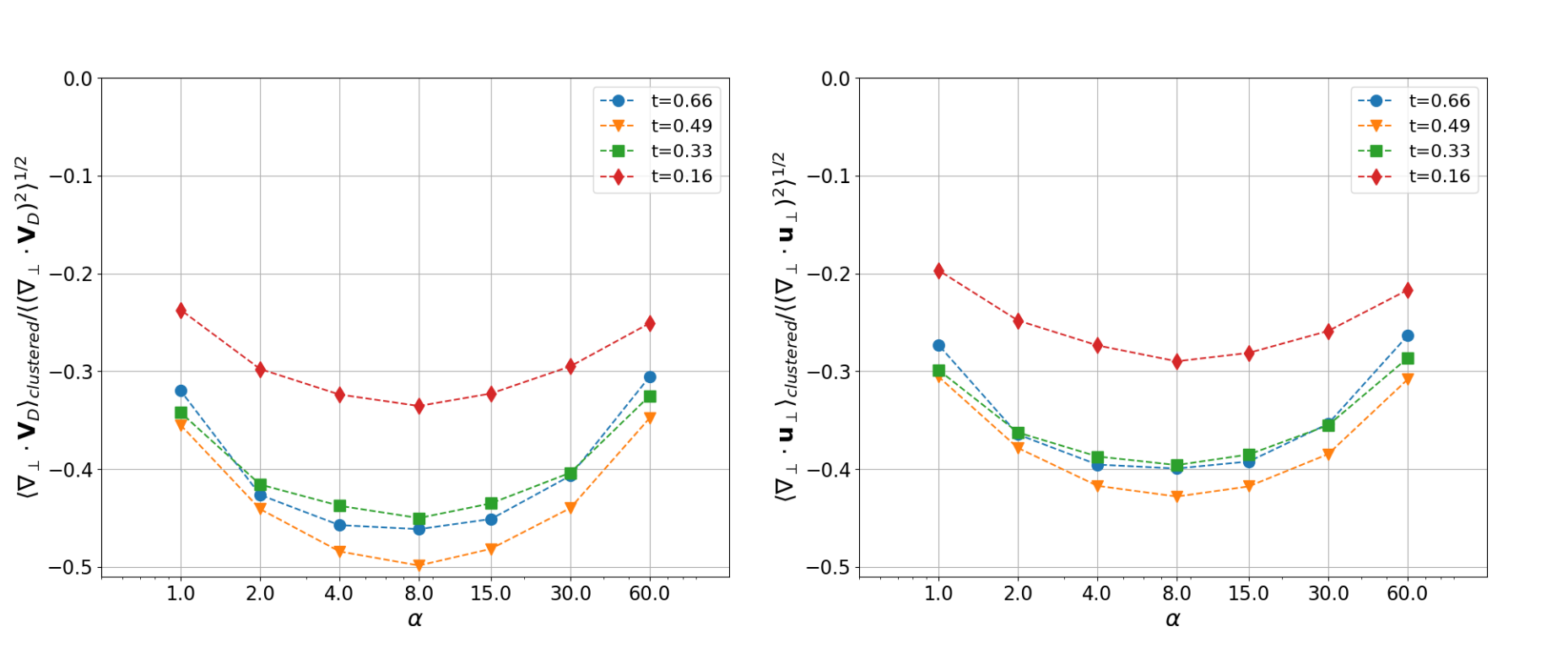}
\caption{As a function of $\alpha$, mean perpendicular divergence of (left) the drift velocity and (right) the velocity field experienced by the clustered particles, normalized by the rms value of the field.}
\label{fig:div_perp_clustered}
\end{figure}

In section \ref{ssec:GCM}, we have shown this model to be applicable to particle motion in the perpendicular plane, so we expect the analogy to hold in 2D.
In the left panel of Figure \ref{fig:div_perp_clustered} we show the mean value of the perpendicular divergence $\nabla_\perp \cdot \mathbf{V}_D = \partial_x V_{D, x} + \partial_y V_{D, y}$ experienced by the particles, for different times and values of $\alpha$.
As the mean divergence for the field is null, we normalized by its standard deviation to show its significance.
It is clear that particles tend to accumulate in regions with negative perpendicular divergence, as expected.
In the right panel of Figure \ref{fig:div_perp_clustered} we show the perpendicular divergence of the MHD velocity field $\mathbf{u}$, which displays an identical behaviour, aside from a small vertical shift.
This is because both fields $\mathbf{V}_D$ and $\mathbf{u}_\perp$ are very similar, which can be seen introducing Eq. (\ref{eq:gen_ohm}) into Eq. (\ref{eq:def_VD}) and expanding
$$ \mathbf{V}_D \approx \mathbf{u}_\perp - \frac{\epsilon}{\rho}\mathbf{j}_\perp + \frac{1}{B}\left[ \frac{\mathbf{j}_\perp}{R_m} - \frac{\epsilon}{\rho}\frac{\nabla_\perp P_e}{\gamma M^2} \right]\times \hat{z} $$
where we have approximated the magnetic field direction $\mathbf{B}/B \approx \hat{z}$.
The right-most term and middle-term are proportional to $1/B\approx 1/B_0$ and $\epsilon = 1/60$, respectively, both of which should be negligible.
The fact that $\langle |\mathbf{V}_D - \mathbf{u}_\perp|^2 \rangle \leq 0.02\langle |\mathbf{V}_D|^2 \rangle^{1/2}\langle |\mathbf{u}_\perp|^2 \rangle^{1/2}$ for all times confirms this observation.
This is relevant because, although particles follow $\mathbf{V}_D$, we can use the simpler $\mathbf{u}_\perp$ as a reasonable approximation.

\begin{figure}[ht]
\centering
\includegraphics[width=0.49\columnwidth]{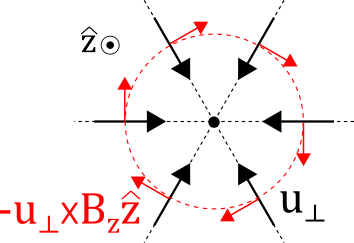}
\includegraphics[width=0.50\columnwidth]{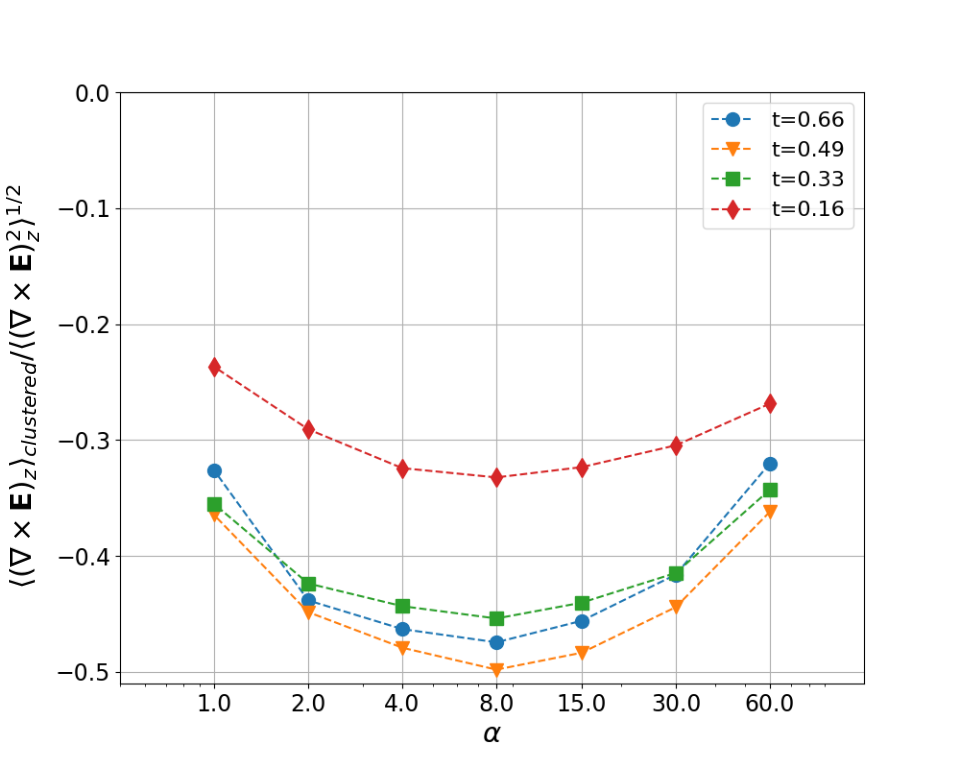}
\caption{(Left) Sketch of the proposed mechanism and (right) mean $z$-component of the electric field curl experienced by the clustered particles, normalized by the rms value of the field.}
\label{fig:curlE}
\end{figure}

This similarity is the bridge to link the clustering with the exceptional perpendicular energization.
In the left panel of Figure \ref{fig:curlE} we illustrate an idealized situation for $\nabla_\perp\cdot\mathbf{u}_\perp < 0$, where streamlines converge to a single point.
This velocity field $\mathbf{u}_\perp$ generates a perpendicular induced electric field $\mathbf{E}_\perp^{ind} = -\mathbf{u}_\perp\times B_z\hat{z}$ that rotates clockwise given that $B_z = B_0+b_z >0$.

This clockwise rotation of the (perpendicular) electric field is equivalent to $(\nabla\times\mathbf{E})_z < 0$, which we show in the right panel of Figure \ref{fig:curlE} to be the case for clustered particles.
In fact, the clear resemblance with Figures \ref{fig:div_perp_clustered} is due to
\[ [\nabla\times\mathbf{E}^{ind}]_z \approx -[\nabla\times(\mathbf{u}\times B_0\hat{z})]_z = B_0\nabla_\perp\cdot\mathbf{u}_\perp \]
Any ion trapped in this region will also gyrate clockwise, resulting in a positive power transfer $\mathcal{P}_\perp = \alpha\mathbf{E}_\perp\cdot\mathbf{v}_\perp > 0$ and energization.
In reality, this alignment will not be perfect, but should be predominant enough to ensure net energization after a full gyroperiod, in a betatron-like resonance (\cite{Swann1933,Dalena2014}).

\begin{figure}[ht]
\centering
\includegraphics[width=0.495\columnwidth, trim={0 2cm 0 2.5cm}, clip]{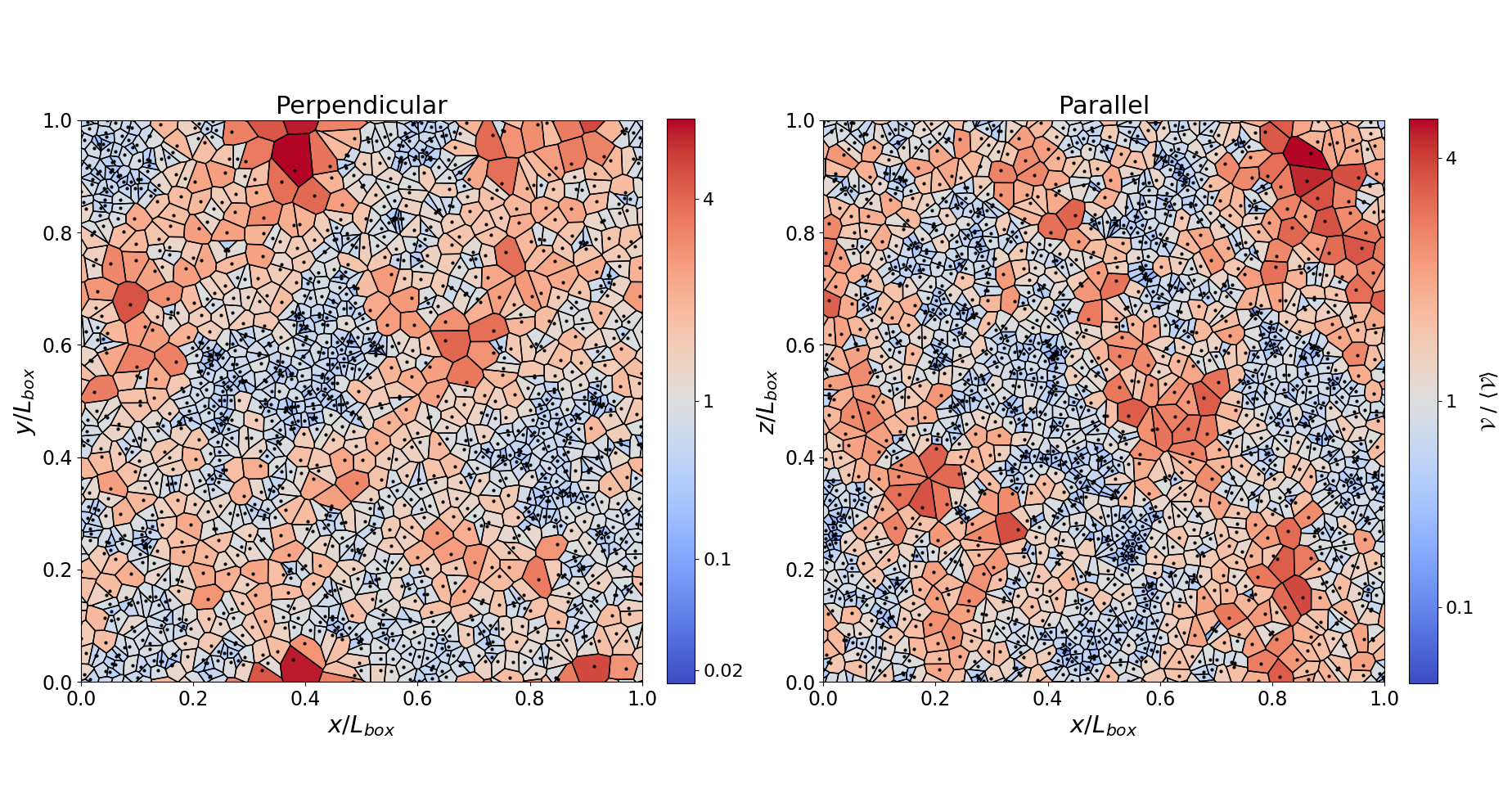} 
\includegraphics[width=0.495\columnwidth, trim={0 2cm 0 2.5cm}, clip]{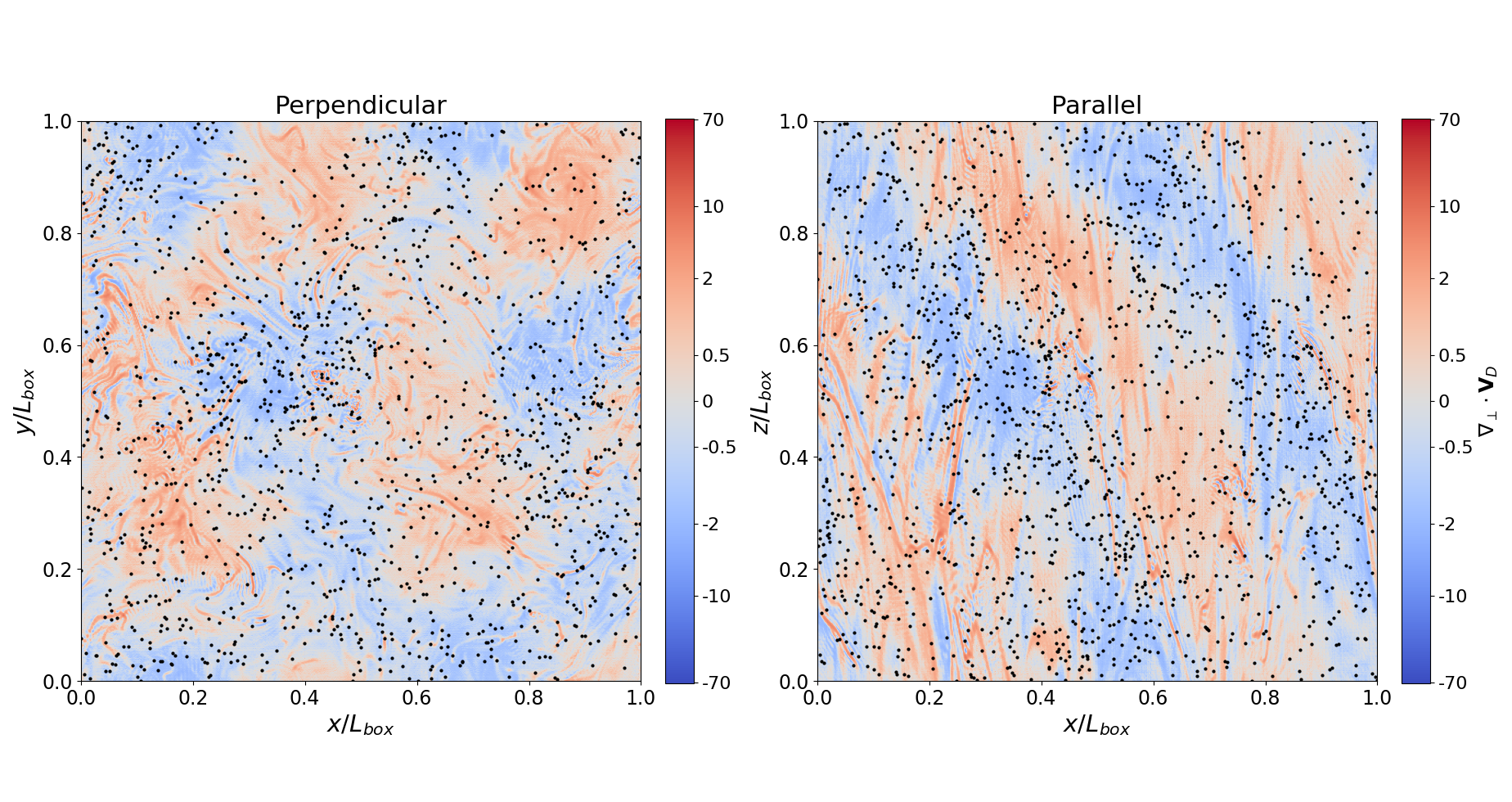}
\includegraphics[width=0.495\columnwidth, trim={0 2cm 0 2.5cm}, clip]{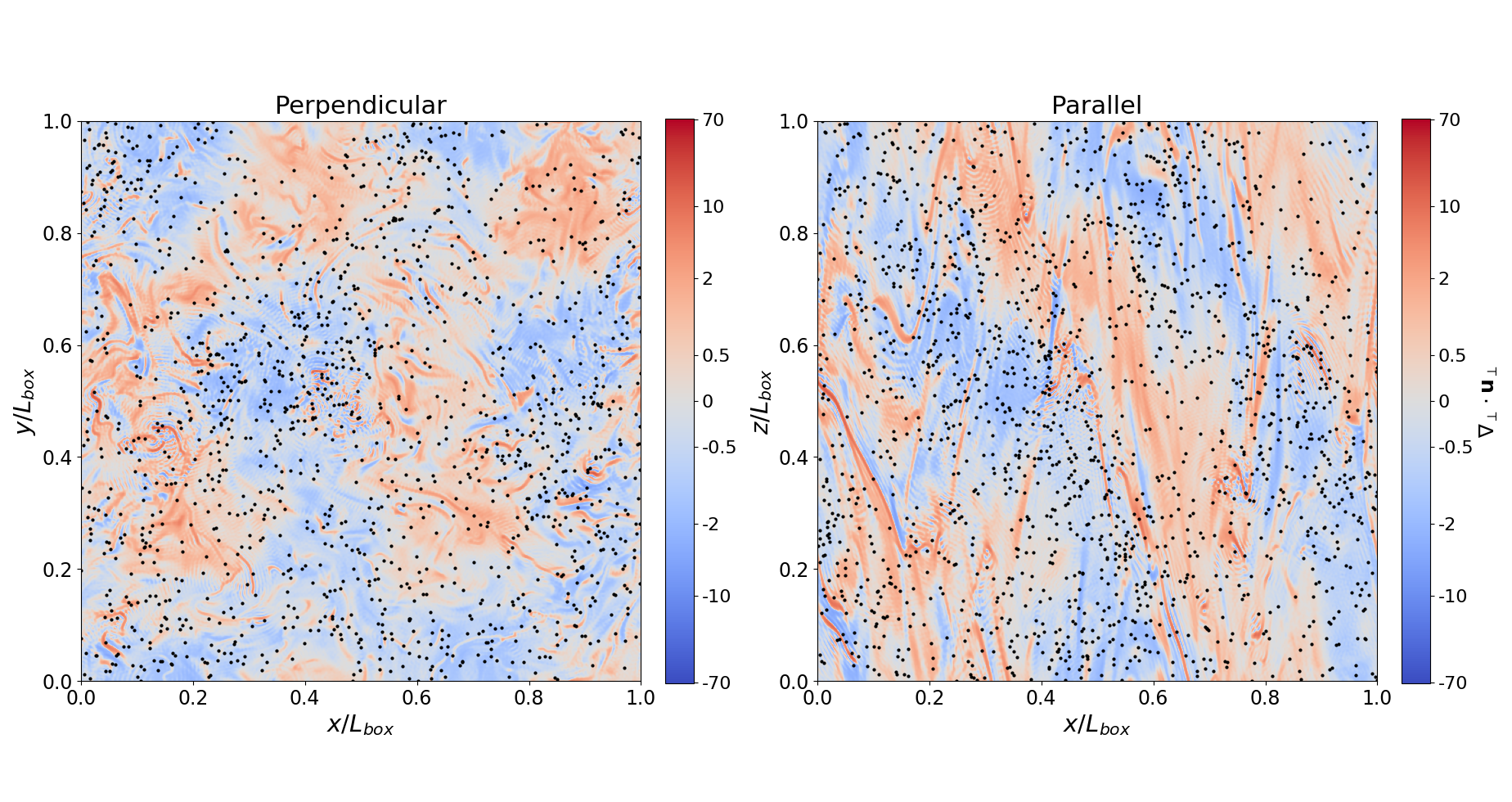}
\includegraphics[width=0.495\columnwidth, trim={0 2cm 0 2.5cm}, clip]{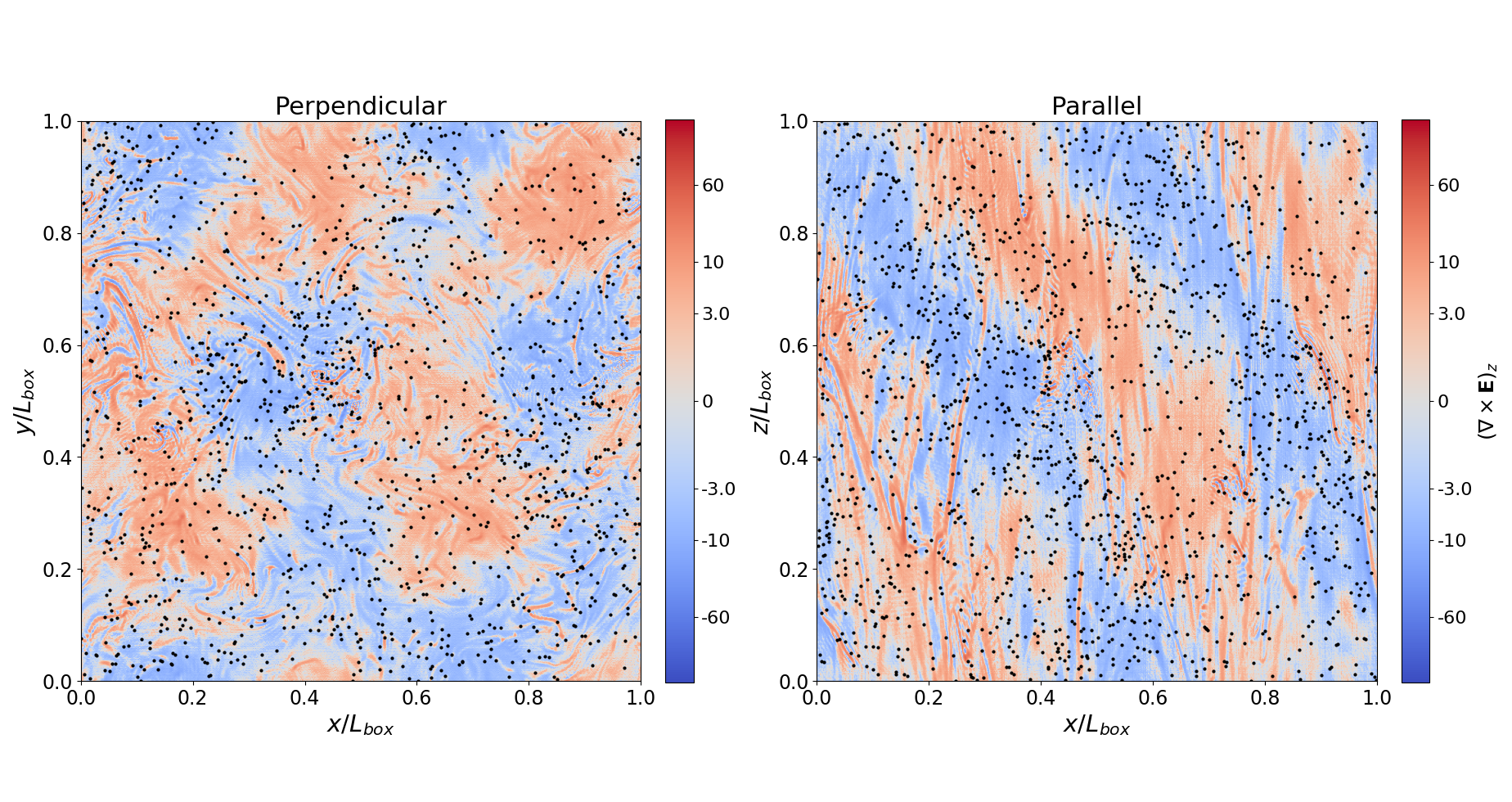}
\caption{For perpendicular and parallel slices; (top left) Voronoi tessellation with cells colored according to their volume (inverse density), (top right) perpendicular divergence of the drift velocity field, (bottom left) perpendicular divergence of the velocity field, and (bottom right) $z$-component of the electric field curl}
\label{fig:comp_voronoi_fields}
\end{figure}

We can clearly visualize the previous analysis in Figure \ref{fig:comp_voronoi_fields}, where the Voronoi tessellation (colored by cell volume) is shown along with the underlying fields $\nabla_\perp\cdot\mathbf{V}_D$, $\nabla_\perp\cdot\mathbf{u}_\perp$ and $(\nabla\times\mathbf{E})_z$ for a perpendicular slice ($z=L_{box}/2$) and a parallel slice ($y=L_{box}/2$) at $t=0.66t_0$ for $\alpha=4$.
The similarity between the 3 (scalar) fields is clear, confirming the previous arguments.
However, more relevant is the correlation between high particle concentration (low cell volume) and the negativity of the fields.
Although this is present in both directions, it is more noticeable in the parallel direction, probably because the structures are more elongated along the $z$-axis.

Nonetheless, this analysis tells us nothing about the parallel component and in fact becomes invalid if particles are not confined in this direction too.
As we have seen in Figure \ref{fig:hist_signs}, parallel motion tends to keep its direction, which implies that only particles with low $|v_z|$ can be long enough in these regions to exploit perpendicular energization.
This is supported by the right panel of Figure \ref{fig:clustered_energ}, but it may not be enough.
A similar analysis for $b_z$ (not shown here) yields that particles tend to cluster in $b_z<0$ regions.
Given that $B_0\gg |\mathbf{b}|$, at first order $|\mathbf{B}| \approx B_0+b_z$ so that particles are clustered in regions of low magnetic field.
This suggests magnetic mirroring as a candidate for parallel confinement and implies low parallel energization to be a \textit{requirement} for clustering, not its \textit{consequence}.

With this model at hand, we are able to make sense of the bell-like behaviour of Figure \ref{fig:sigma_volumes}.
From the last argument, the requirement of low $|v_z|$ is clearly hard to achieve for high $\alpha$ particles, as we see from Figure \ref{fig:energy_rate_vs_alpha} that they tend to increase their parallel energy, thus impairing clustering.
On the other hand, low $\alpha$ particles tend to deviate from the guide center model of Eq. (\ref{eq:def_xD}), as shown by Figure \ref{fig:comp_rmsd_drift} and therefore are not fully trapped by these $\nabla_\perp\cdot\mathbf{u}_\perp < 0$ regions as a true tracer particle.
Another way to view this is to consider that low $\alpha$ particles have higher gyroradius $r_g\sim 1/\alpha$, which increases the chance of them leaving these regions during a gyroperiod.
In any case, these competence is reminiscent of that used at the end of Section \ref{ssec:disp_and_energ} to understand Figure \ref{fig:energy_rate_vs_alpha} and may indeed strengthen it.

To conclude this analysis, we turn back to the unexpected result of the last section: the maxima of the parallel component of Figure \ref{fig:disp_vs_fc}.
Given that low $|v_z|$ is a requirement for clustering, this shows that low parallel displacement may be detrimental for clustering in some cases.
These seemingly contradicting facts could be reconciled by concluding that the initial position of particles in the plasma determines whether they may or may not exploit this mechanism.
Particles that start far away from these regions may never reach them without acquiring some parallel velocity first; this allows them to sample the box but will prevent the trapping from being effective nor lasting long.
These are probably the kind of particles that populate the $f_c\approx 0.2$ region, those that traveled farther to reach this regions and cannot exploit them as much.

\section{Discussion}{\label{sec:disc}}

We studied the dynamics of high-gyroradius (compared to the dissipation scale) or equivalently low charge-to-mass ratio test particles evolving in (weakly) compressible MHD fields.
In terms of energization, the parallel component monotonically increases with $\alpha$ but the perpendicular component displays a maximum around $\alpha\approx4$.
We argued that the energization dependence on $\alpha$  could be understood by considering the effect that $\alpha$ has on the interaction strength and the exploration/exploitation of the MHD structures by the particles.
For the parallel part, low gyroradius (inversely proportional to $\alpha$) allows particles to stick to current channels and exploit its acceleration $\alpha E_z$, so the increase in $\alpha$ both improves the
interaction strength and the confinement in current channels, which favours this parallel energization.  
For the perpendicular component, on the other hand, while the interaction strength increases with $\alpha$, the exploration decreases as particles lower their gyroradius and stick closer to the magnetic field lines.
This is detrimental for perpendicular energization since exploitation requires particle gyroradius to coincide with the characteristic size of certain plasma structures, which amounts to the existence of (at least one) optimal value for $\alpha$.
This is consistent with previews works on the subject (\cite{Dmitruk_2004, LEHE2009}).
By considering the fraction relative to the total energy that corresponds to each component, we removed the interaction strength aspect and confirmed the previous analysis: as $\alpha$ decreases, energization becomes predominantly perpendicular.

For the particles parallel displacement, we found a slightly super-ballistic behaviour, which we explained by noting that most particles move with their original direction along the guide field, experiencing some small acceleration (related to their low parallel energization).
The particles perpendicular displacement displayed a super-diffusive behaviour and mostly independent of $\alpha$.
We applied a guide-center model based on the drift velocity $\mathbf{V}_D=(\mathbf{E}\times \mathbf{B})_\perp/B^2\approx \mathbf{u}_\perp$ which was accurate enough for the rms perpendicular displacement, except for the lowest values of $\alpha$.
This drift velocity had a correlation time $\tau_c\approx t_0/2$, implying some coherence to explain the super-diffusive perpendicular displacement.

We then linked both types of analysis and found that low displacement particles have higher perpendicular energization and lower parallel energization, compared to high displacement particles.
For the parallel energization, this is straightforward as particles with high $v_z^2$ will inevitably travel further, thus increasing their parallel displacement (which is predominant in their total displacement).
For the perpendicular energization, this hinted at a non-trivial connection between perpendicular displacement and energization.
We found this connection through Voronoi analysis, revealing particle clustering in regions with $\nabla_\perp\cdot\mathbf{u}_\perp < 0$, which is reasonable when we consider that particles following the drift velocity $\mathbf{V}_D\approx \mathbf{u}_\perp$ are actually tracers of $\mathbf{u}_\perp$ in the perpendicular plane.
This trapping mechanism requires low initial $v_z$, in order to keep particles from escaping vertically, thus reinforcing the need for low displacement.
In this region, the converging velocity field generates a clockwise circulating electric field ($(\nabla\times \mathbf{E})_z< 0$) where it can be aligned with the cyclotron motion of the particles, imparting net positive energization in each gyroperiod.
This betatron-like resonance (\cite{Swann1933,Dalena2014}) is an efficient mechanism to increase perpendicular energy but it is short lived.
As perpendicular energy increases, pitch-angle scattering may transfer some of it to the parallel component, allowing the particle to escape vertically.
Even if this does not happen, particle gyroradius increases along with the energy and will eventually allow particles to overcome the drift velocity trapping and leave the region.
The importance of $\nabla_\perp\cdot\mathbf{V}_D$ was previously shown in \cite{Lemoine2021}, where a guide center model was also used.

Particles with intermediate $\alpha$ (around $\alpha\approx8$) are most efficiently trapped, as high $\alpha$ particles can more easily acquire the $v_z$ required to escape and low $\alpha$ particles are less subject to $\mathbf{V}_D$ due to their greater gyroradius.
This is an example of the exploitation aspect we discussed at the beginning of this section and is helpful in understanding the bell-like shape found for perpendicular energization.
Although it would be tempting to directly relate this structures to the compressibility of the flow, similar mechanisms are present in incompressible flows (\cite{Boffetta2005, Cressman2007, Stepanov2020}), as long as the dynamics can reasonably be approximated as two-dimensional.
Further work could include the investigation of the effect of compressibility on these structures, their frequency and energization efficiency.
This could partly explain the improvement already observed in particle energization as compressibility increases (\cite{Gonzalez2016}).

In particular, this trapping mechanism is only applicable to particles heavier than protons and is clearly different from those present for electrons (\cite{Ruffolo2003}).
For these low-$\alpha$ particles the drift velocity is comparable to their parallel velocity and fundamental to understand their dynamic.
Although this work is a preliminary exploration, results are consistent with observations, mainly that heavy ions tend to achieve higher thermal velocities than protons (\cite{Hefti1998, Gershman2012, Tracy2015}). The proposed mechanism could help understand this behaviour and chosen values of $\alpha$ can be related to individual ions; N$^+$ or O$^+$ with $\alpha=4$, He$^{2+}$ with $\alpha=30$ and so on.
However, a complete connection with observational results (\cite{Mobius1985, Burlaga1996, Gloeckler1998, Gloeckler2001, Russell2013, Lario2015, McComas2017, Kumar2018}) requires further work.

\begin{acknowledgments}

The authors thank C. Reartes and P.D. Mininni for enlightening exchanges that led to the use of Voronoi analysis for this work.
Support from CONICET and ANPCyT through PIP, Argentina Grant No.~11220150100324CO, and PICT, Argentina Grant No.~2018-4298 is acknowledged. We thank the Physics Department at the University of Buenos Aires for computing time provided in the Dirac cluster.

\end{acknowledgments}

%

\vspace{5mm}






\bibliography{references}
\bibliographystyle{aasjournal}



\end{document}